\documentclass[superscriptaddress,groupedaddress,nofootnoteinbib,12pt]{article} 
\pdfoutput=1
\usepackage{graphicx}
\usepackage{color}
\usepackage{dcolumn}   
\usepackage{bm}       
\usepackage{amssymb}  
\usepackage{amsmath}
\usepackage{sectsty}
\usepackage{latexsym}
\usepackage{float}
\usepackage{ifthen}
\usepackage{caption,subfig}
\usepackage{enumerate}
\usepackage{url}
\usepackage{caption,subfig}
\usepackage{feynmf}	
\usepackage{jcappub}
\usepackage{amsopn}
\usepackage{float}
\usepackage{pst-pdf}

\usepackage{amsfonts}
\usepackage{multirow}
\usepackage{array}
\usepackage{booktabs}
\usepackage{rotating}
\usepackage{color}

\def\({\biggl(}
\def\){\biggr)}
\def\[{\biggl[}
\def\]{\biggr]}
\def\mpl{M_{\rm Pl}}
\def\beq{\begin{equation}}
\def\eeq{\end{equation}}
\def\mn{_{\mu \nu}}

\definecolor{forestgreen}{rgb}{0.133,0.545,0.133}
\newboolean{editorial}
\setboolean{editorial}{true}
\newcommand{\editorial}[2]{\ifthenelse{\boolean{editorial}}{\textcolor{red}{[\textsf{\textbf{{#1}}}: }\textcolor{blue}{\textsf{{#2}}}\textcolor{red}{]}}{}}

\DeclareMathOperator{\tr}{tr}

\renewcommand{\vec}[1]{\bm{\mathrm{{#1}}}}

 \def\be   {\begin{equation}}   \def\ee   {\end{equation}}

 \def\ba  {\begin{eqnarray}}   \def\ea  {\end{eqnarray}}

\renewcommand{\thesubsection}{\arabic{section}.\arabic{subsection}}

\begin{document}

\title{Vainshtein Solutions Without Superluminal Modes}

\author{Gregory Gabadadze$^a$, Rampei Kimura$^a$, David Pirtskhalava$^b$}
\affiliation{
$^a$Center for Cosmology and Particle Physics,
Department of Physics,\\
New York University, New York, NY, 10003}

\affiliation{$^b$Scuola Normale Superiore,\\
Piazza dei Cavalieri 7, 56126, Pisa, Italy}

\vskip 1.9cm

\abstract{
The Vainshtein mechanism suppresses the fifth force at astrophysical distances, while  
enabling it to compete with  gravity  at cosmological scales. Typically,  
Vainshtein solutions exhibit superluminal perturbations.  However, a restricted class of  
solutions with  special boundary conditions were shown to be devoid of  the faster-than-light 
modes.  Here we  extend this class  by finding solutions  in a theory of quasidilaton,  amended by  derivative 
terms consistent with its symmetries.   Solutions with Minkowski  asymptotics are not stable, while   
the ones that exhibit the Vainshtein mechanism by  transitioning  to cosmological backgrounds 
are free of ghosts, tachyons, gradient instability, and  superluminality, 
for all propagating modes present in the theory.  These solutions  
require special choice of the strength and signs of  nonlinear  terms, 
as well as a  choice of asymptotic cosmological boundary conditions. }

\maketitle

\section{Introduction}

Local,  Lorentz-invariant theories that modify gravity at cosmological scales,  ($\sim 10^{28}~cm$), 
do so at the expense of extra gravitationally coupled degrees of freedom. 
The latter need to be "hidden"  at shorter, astrophysical scales 
($ \sim 10^{26}~cm$ and below), to avoid conflict with observations. One of the intricate mechanisms  that provide 
such suppression at shorter scales is the Vainshtein  mechanism  \cite{Vainshtein:1972aa}.
While this mechanism  was originally formulated in the  context of massive gravity, it has 
a broader scope \cite{Deffayet:2002aa,Babichev:2013aa} (for a  nice and comprehensive 
review of screening mechanisms  in cosmology, see \cite{Joyce:2014aa}).  
A majority of  Vainshtein solutions discussed so far 
in the literature exhibit  superluminal perturbations; this has been shown 
in the context of the DGP model \cite{Dvali:2000hr} 
in the decoupling limit \cite{Luty:2003vm}  in Ref. \cite{Nicolis:2004qq,Adams:2006sv}, 
and has been extended to the most general \emph{Galileon} theory in \cite{Nicolis:2008in}. Moreover, it has been demonstrated in \cite{deFromont:2013iwa}, that the same feature persists for Vainshtein solutions in multi-Galileon systems.   Whether or not  superluminalities  always  imply 
acausality -- which is a subtle issue --  will not be discussed here;  instead, we note that 
it  would be easier  if the superluminal modes  were absent altogether. 
An example of a nonlinear scalar theory that exhibits the Vainshtein mechanism without 
superluminal modes  was found  in \cite{Berezhiani:2013aa}. Its features  that enable  
to avoid superluminal modes are:  a choice of a sign of a nonlinear term, and choice of 
boundary cosmological conditions for the solution.

The question is if there are other similar  examples, and if they share  common features with the one 
of \cite{Berezhiani:2013aa}.  In particular  it is interesting to know if such solutions can 
exist in full-fledged theories of massive gravity and their extensions, where tensor and vector 
modes, in addition to scalars,  are also relevant.  These   are  the questions   studied  in the present work.

 Whether or not the graviton can consistently have a nonzero mass has been a subject of discussion for 
more than seven decades. The unique linear theory of a massive spin-2 field, proposed by Fierz and Pauli (FP) \cite{FP:1939aa}, consists of linearized General Relativity (GR) supplemented by a special mass term for the metric perturbation. The special structure of the FP mass guarantees that there are no more than $5$ degrees of freedom propagating on flat space, as required by the representation theory of the Poincar\' e group.
Naively, the massless limit of FP massive gravity 
would be expected to reduce to GR; this however is not the case, and regardless of how small the mass is, the presence of extra degrees of freedom leads to order-one deviations from GR \emph{at all length scales} -- the phenomenon known as the van Dam-Veltman-Zhakarov (VDVZ)
discontinuity \cite{vDV:1970aa,Z:1970aa}.
Continuity in physical predictions can be restored in nonlinear extensions of the the FP gravity through  the {Vainshtein mechanism} \cite{Vainshtein:1972aa}, whereby nonlinear effects screen out extra contributions to the gravitational potential beyond the standard general-relativistic one. The success of the Vainshtein mechanism thus manifestly depends on nonlinear properties of a given extension of the Fierz-Pauli model. 
It has been believed for a long time however,  that  a generic interacting theory of massive gravity 
would necessarily contain  a sixth light ghost degree of freedom -- the so-called Boulware-Deser (BD) ghost \cite{Boulware:1972aa}. The latter would lead  to a catastrophic instability of the system, effectively rendering the Vainshtein mechanism useless. 

This has changed with an explicit construction of the Lagrangian  free from the BD ghost:  First, the BD ghost has been eliminated by a careful choice of the graviton potential consisting of an infinite series of interactions of the form that projects it out order-by-order in perturbation theory \cite{deRham:2010ik}. The infinite series can be resummed into a compact expression \cite{Rham:2011aa}, referred to as the dRGT theory. The full non-perturbative proof of ghost-freedom in this theory has been given in Refs. \cite{Hassan:2011hr,Deffayet:2013aa,Mirbabayi:2011aa,Kugo:2014aa}.

While the dRGT theory guarantees 5  degrees of freedom  on an arbitrary background, 
it only guarantees that these 5 are healthy on  (nearly) flat backgrounds.  This is because
the theory is strongly coupled at a low scale -- incomplete in that sense --  
and some of the 5 modes may flip signs of their  kinetic terms  on strong enough backgrounds, 
converting them into ghost (these latter should not be confused with the BD ghost that is absent now).

Perhaps the minimal way to extend the dRGT model is by introducing into the theory a new scalar field $\sigma$, referred to as quasidilaton, which nonlinearly realizes an Abelian global symmetry \cite{DAmico:2012aa}, 
\begin{eqnarray}
\sigma \to \sigma-\alpha M_{\rm Pl}~, ~~~~~\phi^a \to e^\alpha\phi^a~.
\end{eqnarray}
Here, $\phi^a$ are the four auxiliary (St\"{u}ckelberg) fields, required to formulate the theory in a diffeomorphism-invariant way.
It has been shown recently \cite{Gabadadze:2014aa}, that the quasidilaton  
admits self-accelerated solutions in the decoupling limit, similar to the ones of dRGT gravity \cite{Rham:2011ab}. An important difference from massive gravity however, is that the presence of the quasidilaton makes it possible to avoid all of the stability problems associated with the former class of cosmologies. 

The key aspect for viability of cosmological solutions in modified gravity is the existence and stability of a mechanism that would allow to screen extra contributions to the gravitational potential at distances where GR  agrees 
with observations with an excellent accuracy.
In massive gravity, as discussed above, it is the Vainshtein mechanism that makes this possible. The analysis of spherically symmetric solutions in the decoupling limit of the dRGT theories has revealed however that in general, the Vainshtein mechanism is accompanied with various kinds of instability \cite{Berezhiani:2013ab}. The only way to avoid these is to  
restrict to a particular corner of the parameter space, where the scalar and the tensor modes can be decoupled by a local field redefinition \cite{Berezhiani:2013aa}.  The obtained solution has no superluminal modes.

The above observations motivate to look for a stable realization of the Vainshtein mechanism in quasidilaton theories.  We will focus on the decoupling limit theory, analogous to the one of massive gravity with stable Vainshtein solutions, where the tensor and the scalar modes can be treated independently, but will also account for   the vector modes.
We will show that in a large fraction of the free parameter space the solutions are pathological, 
as they exhibit various instabilities. However, we'll find a small region of the parameter space 
where   a satisfactory solution can be obtained. This solution exhibits the Vainshtein mechanism without 
instabilities  or superluminal modes, and asymptotes to a cosmological  solution  away  from a source.

The rest of this paper is organized as follows: we will start with a brief review of the original quasidilaton theory in Sec. 2, and derive its decoupling limit action, along with the equations of motion for spherically symmetric configurations in Sec. 3. In Secs. 4 and 5, we carry out a detailed analysis of the time-independent and time-dependent solutions respectively. In Sec 6. we consider the decoupling limit action of the most general quasidilaton theory, obtained by supplementing the original Lagrangian by the Horndeski terms for $\sigma$ \cite{Horndeski:1974aa,Deffayet:2013ab,Deffayet:2009aa}. This will provide the first completely stable realization of the Vainshtein mechanism in the given class of theories. 
We summarize our results in Sec 7. 

We adopt the signature $(-,+,+,+)$ for the metric throughout this work, and use the following notation for various contractions of rank-2 tensors: 
$${\cal K}^{\mu}_{~\mu}=[{\cal K}],~~
{\cal K}^{\mu}_{~\nu}{\cal K}^{\nu}_{~\mu}=[{\cal K}^2],~~
{\cal K}^{\mu}_{~\alpha}{\cal K}^{\alpha}_{~\beta}{\cal K}^{\beta}_{~\mu}=[{\cal K}^3]~, \text{~etc.}$$ Moreover, certain expressions involving the Levi-Civita tensor will be shortcut in the following way: 
$\varepsilon^{\mu\alpha\rho\sigma}\varepsilon^{\nu\beta}_{~~\rho\sigma}\Pi_{\mu\nu}\Pi_{\alpha\beta}
\equiv\varepsilon\varepsilon\Pi\Pi$, 
$\varepsilon_\mu ^{~\gamma\alpha\rho}\varepsilon_{\nu \gamma}^{~~\beta\sigma}
\Pi_{\alpha\beta}\Pi_{\rho\sigma}\equiv\varepsilon_\mu\varepsilon_\nu\Pi\Pi$, 
$(B^2)^\mu_\nu\equiv B^\mu_{~\alpha} B^\alpha_{~\nu}$,
$\varepsilon\varepsilon B\partial A \equiv \varepsilon_{\mu_1 \mu_2 \mu_3 \mu_4}\varepsilon^{\nu_1 \nu_2 \mu_3 \mu_4} B^{\mu_1}_{~\nu_1} \partial_{\nu_2} A^{\mu_2}$, and so on.

\section{The Quasidilaton}
The dRGT theory is specified by supplementing the standard Einstein-Hilbert action with special mass and potential terms for the metric perturbation. In its diff-invariant formulation involving four scalar St\" uckelber fields $\phi^a$, the theory takes on the following form \cite{Rham:2011aa}
\begin{eqnarray}
  S_{MG}={M_{\rm Pl}^2\over 2}\int d^4x \sqrt{-g}\left[R
    -{m^2 \over 4}\left( {\cal U}_2+ \alpha_3 {\cal U}_3 + \alpha_4 {\cal U}_4\right)
  \right]+S_m[g_{\mu\nu}, \psi],
\end{eqnarray}
where we have defined
\begin{eqnarray}
\label{Potential}
  {\cal U}_2&=&
4\left([{\cal K}^2]-[{\cal K}]^2\right),
  \nonumber\\
  {\cal U}_3&=&
-[{\cal K}]^3+3[{\cal K}][{\cal K}^2]
  -2[{\cal K}^3],
  \\
  {\cal U}_4&=&
-[{\cal K}]^4+6[{\cal K}]^2[{\cal K}^2]
  -3[{\cal K}^2]^2-8[{\cal K}][{\cal K}^3]+6[{\cal K}^4],\nonumber
\end{eqnarray}
and 
\begin{eqnarray}
  {\cal K}^{\mu}_{~\nu}=\delta^{\mu}_{~\nu}
  -\sqrt{\eta_{ab}g^{\mu\alpha}\partial_{\alpha}\phi^a\partial_{\nu}\phi^b}.
\end{eqnarray}
One can always fix the unitary gauge $\phi^a=\delta^a_\mu x^\mu$, in which all five degrees of freedom, present in the theory sit in the metric perturbation $h_{\mn}\equiv g_{\mn}-\eta_{\mn}$.

It is sometimes useful to view the four scalars $\phi^a$ as certain target-space coordinates of a flat manifold in which our dynamical manifold, parametrized by the coordinates $x^\mu$ is embedded as a spacetime-filling brane. A natural question  is then whether one can define a theory, invariant under \textit{quasidilatations} -- a global Abelian symmetry, under which the target space coordinates scale with respect to those of the dynamical spacetime, $\phi^a\to e^\alpha\phi^a$. This requires introducing a goldstone field $\sigma$ -- \textit{the quasidilaton} -- that nonlinearly realizes the symmetry at hand
\begin{eqnarray}
  \sigma \to \sigma-\alpha M_{\rm Pl}~
\label{Symmetry}
\end{eqnarray}
and enters the action through an extended ${\cal K}$ - tensor  
\begin{eqnarray}
\bar{{\cal K}}^{\mu}_{~\nu}=\delta^{\mu}_{~\nu}-e^{\sigma/ M_{\rm Pl}}\sqrt{\eta_{ab}g^{\mu\alpha}\partial_{\alpha}\phi^a\partial_{\nu}\phi^b}~.
\end{eqnarray}
Then the full action including the quasidilaton is given by the following expression
\cite{DAmico:2012aa}
\begin{eqnarray}
S&=&{M_{\rm Pl}^2\over 2}\int d^4x \sqrt{-g}\left[R
-{\omega}g^{\mu\nu}\partial_{\mu}\sigma\partial_{\nu}\sigma
-{m^2 \over 4}\left( {\bar{\cal U}}_2+ \alpha_3 {\bar{\cal U}}_3 + \alpha_4 {\bar{\cal U}}_4\right)
\right]\nonumber\\
&+&S'+S_m[g_{\mu\nu}, \psi]+\dots ~.
\label{Action}
\end{eqnarray}
Here we have added a kinetic term for the new scalar $\sigma$, and 
the potentials ${\bar{\cal U}}_i$ are defined in terms of $\bar{{\cal K}}$ as in  (\ref{Potential}). 
In addition, we have allowed for an extra piece in the action, invariant under the quasidilaton symmetry (\ref{Symmetry}),
\begin{eqnarray}
S'= M_{\rm Pl}^2m^2 \alpha_5 \int d^4x \sqrt{-g} \,
e^{4\sigma / M_{\rm Pl}} \sqrt{\det{(g^{\mu\alpha}\partial_\alpha\phi^a\partial_\nu\phi_a)}}.
\label{Action2}
\end{eqnarray}
In the dRGT theory, 
this term is non-dynamical, ${\cal L}' \sim \sqrt{-\eta}$,
which is however not true in the presence of the quasidilaton. Moreover, it includes a tadpole for $\sigma$ and is therefore expected to contribute to asymptotically non-trivial backgrounds, which we will be interested in in this paper. Furthermore, the ellipses denote possible extra terms involving $\sigma$ consistent with the quasidilaton symmetry, that we will consider in what follows.

A further extension of the quasidilaton has been found in Ref. \cite{Felice:2013aa}, obtainable via replacing $f_{\mu\nu}$ with a new fiducial metric
\beq
{\bar f}_{\mu\nu} \equiv f_{\mu\nu}- (\alpha_\sigma /\mpl^2m^2) {\rm e}^{-2\sigma/\mpl}\partial_\mu \sigma\partial_\nu \sigma~.
\eeq
The resultant theory is still manifestly invariant under (\ref{Symmetry}) and with a little more work one can show that it is also devoid of the BD ghost (see \cite{Mukohyama:2013aa} for a detailed discussion). 

We will focus on the action (\ref{Action}) - (\ref{Action2}) for definiteness throughout the present paper. In fact, the theory we consider leads to the \textit{most general decoupling limit action of a tensor and two scalars, invariant under galilean symmetry}. Since we are primarily interested in the decoupling limit in this work, we expect our analysis to capture the phenomenological aspects of any extension of massive general relativity, based on the quasidilaton and (approximate) galilean symmetries.

\section{Decoupling limit}
\label{sec:3}
In gauge theories in general, and in massive gravity in particular, there exists a very convenient regime of the theory -- \emph{the decoupling limit} -- where most of the complications associated with the low-energy dynamics go away. In the case of the ghost-free massive general relativity, the decoupling limit captures physics at distances in the range $(\mpl m^2)^{-1/3}< r < m^{-1}$, essentially encompassing all relevant astrophysical and cosmological scales (given that the most reasonable choice for the graviton mass is around the current Hubble scale $m\sim H_0^{-1}$). 
Our analysis of spherically symmetric solutions in theories with the quasidilaton will be carried out exclusively in this limit.

Let us consider small fluctuations of the St\"{u}ckelberg fields around their unitary gauge values,
\begin{eqnarray}
  \phi^a=\delta^a_\mu x^\mu - {\eta^{a\mu} A_\mu  \over  M_{\rm Pl}m}
- {\eta^{a\mu} \partial_\mu \pi \over M_{\rm Pl}m^2},
\end{eqnarray}
while the metric is expanded around the Minkowski spacetime in the usual way,
$g_{\mu\nu}=\eta_{\mu\nu}+h_{\mu\nu}/\mpl$.
The decoupling limit, in which $\pi, A$ and $h$ capture respectively the helicity- 0, 1, and 2 components of the massive graviton, is then defined in the following way
\begin{eqnarray}
  M_{\rm Pl} \to \infty, \qquad m \to 0, 
  \qquad \Lambda = (M_{\rm Pl}m^2)^{1/3}={\rm fixed}, 
  \qquad {T_{\mu\nu} \over M_{\rm Pl}}={\rm fixed}~,
\end{eqnarray}
and the scalar-tensor part of the action for massive GR plus the quasidilaton reduces to the following expression
\begin{eqnarray}
  &&{\cal L}_{\rm DL}^{( h,\pi,\sigma)}
=-{1\over 4} h^{\mu\nu}{\cal E}_{\mu\nu}^{\alpha\beta}h_{\alpha\beta}
  -{\omega \over 2} \partial^\mu\sigma\partial_\mu\sigma
  -h^{\mu\nu} \left[{1\over 4}\varepsilon_\mu\varepsilon_\nu \Pi 
    -{\alpha \over 4 \Lambda^3}\varepsilon_\mu\varepsilon_\nu \Pi\Pi
    -{\beta\over 2\Lambda^6}\varepsilon_\mu\varepsilon_\nu \Pi\Pi\Pi\right]\nonumber\\
  &&~~~~~~~~~~~
+\sigma\left[4\alpha_5 \Lambda^3+ \gamma_0\varepsilon\varepsilon\Pi
    +{\gamma_1\over \Lambda^3}\varepsilon\varepsilon\Pi\Pi
    +{\gamma_2\over \Lambda^6}\varepsilon\varepsilon\Pi\Pi\Pi
    +{\gamma_3\over \Lambda^9}\varepsilon\varepsilon\Pi\Pi\Pi\Pi\right]\nonumber\\
  &&~~~~~~~~~~~
  +{1 \over 2M_{\rm Pl}}h^{\mu\nu}T_{\mu\nu} ~,
\label{LagrangianDL}
\end{eqnarray}
where we have made use of the following notation,
\begin{eqnarray}
  &&\alpha=-{3 \over 4 }\alpha_3 -1,\qquad
  \beta=-{1 \over 8} \alpha_3 - {1\over 2}\alpha_4,\qquad
  \gamma_0={1 \over 2} -{2 \over 3} \alpha_5, \nonumber\\
  &&\gamma_1={3 \over 8} \alpha_3 - {1\over 2}-\alpha_5,\qquad
  \gamma_2={1 \over 2} \alpha_4 - {3\over 8}\alpha_3 - {2\over 3}\alpha_5,\qquad
  \gamma_3=-{1 \over 2} \alpha_4 - {1\over 6}\alpha_5~.
\end{eqnarray}
The lagrangian ${\cal L}_{\rm DL}^{( h,\pi,\sigma)}$ is invariant under linearized gauge transformations $h_{\mu\nu} \to h_{\mu\nu} + \partial_{\mu}\xi_{\nu}+\partial_{\nu}\xi_{\mu}$, as well as internal galilean transformations for $\pi$ and $\sigma$, $\partial_\mu \pi \to \partial_\mu \pi + c_\mu$ and $\partial_\mu \sigma \to \partial_\mu \sigma + d_\mu$.
Furthermore, the complete decoupling limit additionally features the mixing and interaction terms for the gelicity-1 and helicity-0 gravitons specified by the following action
\begin{eqnarray}
  &&{\cal L}_{\rm DL}^{(A)}=-{1\over 4} 
\biggl[
\Lambda^3 \varepsilon\varepsilon BB + 2(1-\alpha)\varepsilon\varepsilon BB \Pi
-{\alpha+6\beta \over \Lambda^3}\varepsilon\varepsilon BB\Pi\Pi
+\varepsilon\varepsilon B^2 \Pi
 -{ \alpha \over \Lambda^3} \varepsilon\varepsilon B^2 \Pi\Pi
\nonumber\\
  &&~~~~~~~~~~~
-{2 \beta \over \Lambda^3}\varepsilon\varepsilon B^2 \Pi\Pi\Pi
+2\Lambda^{3/2} \varepsilon\varepsilon B \partial A 
-{4 \alpha \over \Lambda^{3/2}}\varepsilon\varepsilon B \partial A \Pi 
-{12 \beta \over \Lambda^{9/2}}\varepsilon\varepsilon B \partial A \Pi \Pi
\biggr].
\label{LagrangianVector}
\end{eqnarray}
Here, $B_{\mu\nu}$ is an auxiliary non-dynamical anti-symmetric tensor which can in principle be algebraically integrated out\footnote{The vector-scalar lagrangian \eqref{LagrangianVector} has first been derived in the vielbein formalism in \cite{Gabadadze:2013aa,Ondo:2013wka}.}, and the action is invariant under $U(1)$ gauge transformations, $A_\mu \to A_\mu + \partial_\mu \chi$. For the spherically symmetric solutions we consider below, $A_\mu = 0$ at the background level and the vector action will only be relevant for their perturbative stability.

In the special case of $\beta=0$, the helicity-2 mode in (\ref{LagrangianDL}) can be completely decoupled from the rest of the fields through the following field redefinition
\begin{eqnarray}
\label{newframe}
  h_{\mu\nu} \to h_{\mu\nu}+\pi\eta_{\mu\nu}
  -{\alpha \over \Lambda^3} \pi \Pi_{\mu\nu},
\end{eqnarray}
which is not true in the presence of the $h^{\mu\nu}\varepsilon_\mu\varepsilon_\nu \Pi\Pi\Pi$ interaction. The quasidilaton action in the new frame defined by \eqref{newframe} is then given by the following bi-\emph{Galileon} \cite{Nicolis:2008in} theory 
\begin{eqnarray}
  &&{\cal L}_{\rm DL}=-{1\over 4} h^{\mu\nu}{\cal E}_{\mu\nu}^{\alpha\beta}h_{\alpha\beta}
  -{\omega \over 2} \partial^\mu\sigma\partial_\mu\sigma
  \nonumber\\
  &&~~~~~~~~~~~
  -{1\over 8} \pi\left[\varepsilon\varepsilon \Pi 
    -{2 \alpha \over \Lambda^3}\varepsilon\varepsilon \Pi\Pi
    +{\alpha^2 - 4 \beta \over \Lambda^6}
    \varepsilon\varepsilon\Pi\Pi\Pi
    +{4\alpha\beta \over \Lambda^9}\varepsilon\varepsilon \Pi\Pi\Pi\Pi
  \right]
  \nonumber\\
  &&~~~~~~~~~~~
+\sigma\left[4\alpha_5 \Lambda^3+ \gamma_0\varepsilon\varepsilon\Pi
    +{\gamma_1\over \Lambda^3}\varepsilon\varepsilon\Pi\Pi
    +{\gamma_2\over \Lambda^6}\varepsilon\varepsilon\Pi\Pi\Pi
    +{\gamma_3\over \Lambda^9}\varepsilon\varepsilon\Pi\Pi\Pi\Pi\right]
  \nonumber\\
  &&~~~~~~~~~~~
    +{1 \over 2M_{\rm Pl}}h^{\mu\nu}T_{\mu\nu}
    +{1 \over 2M_{\rm Pl}}\pi T
    -{\alpha \over 2M_{\rm Pl}\Lambda^3} \pi \Pi_{\mu\nu}T^{\mu\nu}.
   \label{LagrangianDL2}
\end{eqnarray}
Throughout this paper we will concentrate on $\beta=0$ for simplicity.
Varying the action with respect to $\pi$, one obtains
\begin{eqnarray}
&&{1\over 4}\varepsilon\varepsilon \Pi 
-{3\alpha \over 4 \Lambda^3}\varepsilon\varepsilon \Pi\Pi
+{\alpha^2\over 2\Lambda^6}\varepsilon\varepsilon \Pi\Pi\Pi
-\gamma_0\varepsilon\varepsilon\Sigma
-{2\gamma_1\over \Lambda^3}\varepsilon\varepsilon\Sigma\Pi
-{3\gamma_2\over \Lambda^6}\varepsilon\varepsilon\Sigma\Pi\Pi\\
&&~~~~~~~~~~~~~~~~~~~~~~~~~~~~
-{4\gamma_3\over \Lambda^9}\varepsilon\varepsilon\Sigma\Pi\Pi\Pi
={1 \over 2M_{\rm Pl}}T-{\alpha \over M_{\rm Pl}\Lambda^3}\Pi_{\mu\nu}T^{\mu\nu}~,
\end{eqnarray}
while the $\sigma$-equation of motion reads
\begin{eqnarray}
&&-{\omega \over 6} \varepsilon\varepsilon\Sigma
+4\alpha_5 \Lambda^3+\gamma_0\varepsilon\varepsilon\Pi
    +{\gamma_1\over \Lambda^3}\varepsilon\varepsilon\Pi\Pi
    +{\gamma_2\over \Lambda^6}\varepsilon\varepsilon\Pi\Pi\Pi
    +{\gamma_3\over \Lambda^9}\varepsilon\varepsilon\Pi\Pi\Pi\Pi=0~.
\end{eqnarray}
The tensor mode on the other hand obeys exactly the same Einstein's equations as in general relativity.
In what follows, we will allow for time-dependent background solutions for the scalars $\pi$ and $\sigma$. To this end, the general ansatz that we will adopt has the following form
\begin{eqnarray}
  &&\pi(t,x) \to {a \over 2} \Lambda^3 t^2 + \pi(r),\nonumber\\
  &&\sigma(t,x) \to {b \over 2} \Lambda^3 t^2 + \sigma(r)~,
\label{backgroundPiSigma}
\end{eqnarray}
which reduces the $\pi$-equation of motion to 
\begin{eqnarray}
  && -{a \over 2} + 2 b \gamma_0
+\left({3 \over 2} + 3 a \alpha + 4b \gamma_1\right)\lambda
-(3\alpha+3a\alpha^2-6b\gamma_2)\lambda^2
+(\alpha^2+8b\gamma_3)\lambda^3
\nonumber\\
&&  
-(6\gamma_0-4a\gamma_1)\lambda_\sigma
-(8\gamma_1-12a \gamma_2)\lambda_\sigma \lambda
-(6\gamma_2 - 24a \gamma_3)\lambda_\sigma \lambda^2
=(1+2a\alpha)\left({r_* \over r} \right)^3~,
  \label{EOM_pi}
\end{eqnarray}
while the $\sigma$ e.o.m yields
\begin{eqnarray}
&&4\alpha_5 + 6 a \gamma_0 - b \omega
  -6(3\gamma_0-2a\gamma_1)\lambda
\nonumber\\
&&  ~~~~~~~~~~
  -6(2\gamma_1-3a\gamma_2)\lambda^2
  -6(\gamma_2-4 a \gamma_3)\lambda^3
  +3\omega\lambda_\sigma
=0~.
  \label{EOM_sigma}
\end{eqnarray}
Here, we defined the dimensionless variables $\lambda$, $\lambda_{\sigma}$ 
and the Vainshtein radius $r_*$ as follows,
\begin{eqnarray}
  &&\lambda\equiv {\pi' \over \Lambda^3 r}, \qquad 
  \lambda_\sigma\equiv {\sigma' \over \Lambda^3 r}, \qquad
  r_*\equiv \left({M\over 4\pi M_{\rm Pl}^2 m^2}\right)^{1/3}.
\end{eqnarray}

\section{Time-independent background solutions}
\label{sec:4}
In this section, we assume a static ($a=b=0$) background solution and study its stability under small perturbations. Moreover, we will set $\alpha_5=0$ to start with.
In this case, the equation of motion for $\lambda$ can be 
obtained by combining Eqs. (\ref{EOM_pi}) and (\ref{EOM_sigma}),
\begin{eqnarray}
  &&{3 \over 2} \left(1-{6 \over \omega}\right) \lambda
  -\left(3\alpha + {36 \gamma_1 \over \omega}\right) \lambda^2 
  + \left(\alpha^2-{32 \gamma_1^2 + 24\gamma_2 \over \omega}\right) \lambda^3\nonumber\\
  &&~~~~~~~~~~~~~~~~~~~~~~~~~~~~~~~~~~~~~~~~~~~~~~
  -40 {\gamma_1\gamma_2 \over \omega}\lambda^4 
  -{12\gamma_2^2 \over \omega}\lambda^5 
  = \left({r_* \over r}\right)^3.
\label{master}
\end{eqnarray}
For $r \gg r_*$, five possible asymptotic solutions are
obtained by solving $P(\lambda)=0$, the latter function defined by the left hand side of Eq.~(\ref{master}).
One is the trivial $\lambda=0$, and the rest of these we denote by $\lambda_{1,2,3,4}=const$.
The solution $\lambda(r\to \infty)=0$ corresponds to asymptotically Minkowski geometry, the leading piece given as follows
\begin{eqnarray}
  \lambda \simeq \frac{2 \omega}{3(\omega-6)}\left(\frac{r_*}{r}\right)^3~.
  \label{Minkowsol}
\end{eqnarray}
The other $\lambda(r\to\infty)\neq 0$ solutions correspond to cosmological backgrounds. Inside the Vainshtein radius, $r \ll r_*$, the highest nonlinear term $\lambda^5$ dominates, 
and there are two solutions depending on the sign of $\omega$, 
\begin{eqnarray}
  \lambda \simeq \pm\left({ 3 |\omega|\over 16(1+\alpha)^2}\right)^{1/5} 
  \left({r_* \over r}\right)^{3/5}.
\label{solin}
\end{eqnarray}
Here, negative $\lambda$ corresponds to positive $\omega$, and vice versa.
As it can be straightforwardly verified from the explicit form of the action, negative $\omega$ unambiguously leads to a ghost in the $\sigma$ field, so we disregard this possibility, fixing the lower sign in (\ref{solin}).
This solution is the only one inside the Vainshtein radius -- no matter what it matches to outside.

It is interesting to evaluate the effective energy density and pressure
contributed from the scalar modes, i.e. effective $\pi$ stress tensor, to which the physical metric (the one before the field redefinition \eqref{newframe}) couples to. 
For the asymptotic Minkowski solution, the effective energy density at large distances reads
\begin{eqnarray}
  \rho=M_{\rm Pl}G_{00}\simeq-\frac{4\alpha\omega^2\Lambda^3 M_{\rm Pl}}{3(w-6)^2}
  \left(\frac{r_*}{r}\right)^6,
\end{eqnarray}
while the pressure is given by the following expression
\begin{eqnarray}
  p={M_{\rm Pl} \over 3} G^i_{i}
\simeq \frac{4\alpha\omega^2\Lambda^3 M_{\rm Pl}}{9(w-6)^2}
  \left(\frac{r_*}{r}\right)^6~,
\end{eqnarray}
rendering the effective equation of state $w\simeq -1/3$. 
Positive energy density requires $\alpha<0$
for this solution; negative $\alpha$ however will always lead to a ghost inside reasonable sources \footnote{
One can see this by e.g. considering a static lump of dust of constant density,
$T_{\mu\nu}=\rho \delta_\mu^0\delta_\nu^0 \theta(R-r)$,
where $R$ denotes its size.
Then, the leading contribution to the kinetic term for the $\pi$-perturbations is given by
$(\alpha \rho /M_{\rm Pl} \Lambda^3)(\delta\dot{\pi})^2$,
leading to a ghost unless $\alpha>0$.
Now, the structure of the matter couplings to gravity in the present case is the same as in the original dRGT theory, so that the same argument goes through here to constrain the sign of $\alpha$. 
} \cite{Berezhiani:2013aa}.
One thus concludes that the asymptotically Minkowski solutions are always plagued by a ghost instability.
For the rest of the constant $\lambda$ solutions, $\lambda_{1,2,3,4}$,
the asymptotic effective energy density and pressure read
\begin{eqnarray}
  \rho\simeq-3\lambda(1-\alpha\lambda)\Lambda^3 M_{\rm Pl},\qquad p \simeq-\lambda(-2+\alpha\lambda)\Lambda^3 M_{\rm Pl}~.
\end{eqnarray}
Since $\lambda$ has to be negative everywhere\footnote{
We have argued below eq. (\ref{solin}), that $\lambda$ has to be negative
inside the Vainshtein radius in order to avoid ghosts.
Now, $P(\lambda)$ becomes infinity as r goes to zero and is everywhere nonzero except for spatial infinity. 
This means that $\lambda$ does not cross zero anywhere in space for the solutions at hand (since $P(0)$ would vanish at a finite distance from the origin if this were not true), being negative also outside of the Vainshtein radius.
}, we have positive energy and negative pressure for the case that the system is ghost free in the region within the source, $\alpha>0$. Whether or not ghost-freedom persists for the rest of the space, we investigate next.

Let us slightly perturb our background solution. The detailed derivation of the perturbation action is given in Appendix \ref{sec:scalar-perturbations}.
The leading piece in the kinetic term for $\pi$-perturbations (denoted by $\phi(t,\vec{x})$ in what follows) in the region inside the Vainshtein radius is given by
\begin{eqnarray}
  {\cal L}^{(2)}_{DL} = -\frac{3^{4/5}}{10\times 2^{1/5}} 
  \frac{(1+\alpha)(10+7\alpha)}{\omega}\bigg [\frac{\omega}{(1+\alpha)^2}\bigg ]^{4/5}
  \left({r_*\over r}\right)^{12/5} \dot\phi ^2+\dots~.
\end{eqnarray}
Since $\alpha$ has to be positive, 
one can see that $\phi$ becoming a ghost somewhere in space is unavoidable.
Let us for completeness also check the kinetic term for the $\sigma$ perturbation (denoted by $\psi(t,\vec{x})$ throughout the present work).
The general expression for the kinetic terms in the quadratic perturbation lagrangian is of the following form
\begin{eqnarray}
  &&{\cal L}^{(2)}_{\rm DL} \supset 
{\cal A}_1 (\partial_t \phi)^2+{\cal B}_1 (\partial_t \psi)^2
+{\cal C}_1 (\partial_t \phi)(\partial_t \psi) 
={\cal A}_1 \left(\partial_t \phi+\frac{{\cal C}_1 }{2{\cal A}_1 }\partial_t \psi\right)^2
+\left({\cal B}_1-\frac{{\cal C}_1^2}{4{\cal A}_1 }\right) (\partial_t \psi)^2.\nonumber
\end{eqnarray}
On the solution \eqref{solin} inside Vainshtein radius, the analysis of Appendix \ref{sec:scalar-perturbations} gives
\begin{eqnarray}
{\cal A}_1 \propto \left({r_* \over r}\right)^{12/5}, \qquad 
{\cal C}_1 \propto \left({r_* \over r}\right)^{9/5}~,
\label{coefficientV}
\end{eqnarray}
meaning that ${\cal C}_1^2/4{\cal A}_1 \propto (r_*/r)^{6/5}\gg {\cal B}_1$ (since ${\cal B}_1=\omega/2$ is just a constant).
We thus arrive at a conclusion that both of the scalar modes are ghosts inside the Vainshtein radius (and outside the source). 
One can show, that including the extra term (\ref{Action2}) in the action does not help: even if we include this term, the qualitative structure of the solutions remains intact. In particular, the coefficients ${\cal A}_1$ and ${\cal C}_1$ 
still go as $({r_* / r})^{12/5}$ and $({r_* / r})^{9/5}$
respectively for $r\ll r_*$, turning at least one of the two scalars into a ghost
on the time-independent solutions.
In the next section, we will attempt to fix the problem by allowing time-dependence for the background.

\section{Time-dependent case}
\label{sec5}
In this section we investigate the case of time-dependent background configurations,
\beq
a\neq 0, \qquad b \neq 0~.
\eeq
We note that while the fields are time dependent, they enter the Lagrangian with derivatives
so that the stress-tensors  of these  fields on the solutions at hand 
are time-independent. In subsections 1 and 2 we still set the parameter $\alpha_5$ to zero, and study 
the case of $\alpha_5 \neq 0$ in subsection 3.

\subsection{Asymptotically de Sitter}
To start with, we consider a solution, corresponding to $1+2\alpha a \neq 0$.
As it can be straightforwardly verified, we then have exactly the same ghost problem inside the Vainshtein radius as described in the end of the previous section\footnote{Indeed, the fact that the background configurations have an additional time-dependent piece can not change the kinetic terms for their perturbations -- they can only affect the gradient energy. This follows from the special galileon structure of our decoupling limit lagrangian.}. Let us nevertheless have a closer look at possible asymptotically de Sitter backgrounds. 
The condition for de Sitter asymptotics can be recast in terms of the effective $r\to \infty$ equation-of-state parameter,
\begin{eqnarray}
  w\equiv {p \over \rho}
=\frac{a-2a\alpha\lambda-\lambda(2-\alpha\lambda)}{3\lambda(1-\alpha \lambda)}=-1~.
\end{eqnarray}
This is solved for $\lambda=1/2\alpha$ and $\lambda = -a$ and both of these conditions can not be imposed at the same time unless $1+2\alpha a$ vanishes.

Let us first consider the case $\lambda=1/2\alpha$. As shown in Appendix \ref{sec:vector-perturbations}, this parameter choice kills the kinetic term for the vector mode at the quadratic level, leading to infinitely strongly coupled vector perturbations. 
The second, $\lambda = -a$ case on the other hand, corresponds to a Lorentz-invariant profile for the $\pi$ field, and has been considered as a special $\beta=0$ subclass of the self-accelerating solutions, found in \cite{Gabadadze:2014aa}\footnote{Note that the parameter space on Fig. 1 of \cite{Gabadadze:2014aa} corresponds to a particular choice of the parameter $\omega$.}. 
These solutions can be discarded on the basis of our analysis of the previous section. Indeed, the Vainshtein solution in the vicinity of the source
\begin{eqnarray}
	\lambda \sim
	\left({r_* \over r}\right)^{3/5},
\end{eqnarray} 
is the same one as in the previous section (with a difference only in numerical coefficients), and therefore at least one of the scalars has to be a ghost inside the Vainshtein radius for our choice of the parameters.

Finally, we look at the case that both conditions, $\lambda=1/2\alpha$ and $\lambda = -a$, are imposed. This is only possible if $1+2\alpha a = 0$, which means that the source term in the scalar background equations vanishes. As already noted above, the condition $\lambda=1/2\alpha$ leads to infinitely strongly coupled vector perturbations, so we discard this possibility. 

\subsection{Solutions with decoupled sources}

We now turn to the special case, $1+2\alpha a = 0$, $\lambda \neq 1/2 \alpha$, and $\lambda \neq -a$, for which the equation of motion for $\pi$ no longer depends on the source. This can potentially take care of the ghost problem inside the Vainshtein radius, since the scalar profiles correspond to $\lambda,\lambda_\sigma=const$ everywhere in space, describing cosmologies with the equation of state parameter $w\neq -1$. 
To simplify the analysis, let us consider the limit $\omega \to \infty$, where the interactions between $\pi$ and $\sigma$ is absent, and expand the solution around it.
To do so, it is convenient to define a canonically normalized $\sigma$ field, $\sigma \equiv {\tilde \sigma / \sqrt{\omega}}$, and consider an expansion in the inverse powers of the large parameter 
\begin{eqnarray}
	\lambda=\lambda_1+ \lambda_2 \omega^{-1/2} +{\cal O}(\omega^{-1}), \qquad
	{\tilde \lambda}_{\sigma}={\tilde \lambda}_{\sigma,1}
	+ {\tilde \lambda}_{\sigma, 2} \omega^{-1/2} +{\cal O}(\omega^{-1})~,
	\label{expandedsigma}
\end{eqnarray}
where  ${ \lambda}_{\sigma}=\tilde \lambda_\sigma/\sqrt{\omega}$. The zeroth order, $\omega\to\infty$ solutions are given as follows
\begin{eqnarray}
	\lambda_1={1 \over  2 \alpha}, \frac{1\pm \sqrt{3}}{2\alpha}, \qquad 
	{\tilde \lambda}_{\sigma,1}={{\tilde b}\over 3},
	\label{lambda1}
\end{eqnarray}
where we have defined ${ b}=\tilde b/\sqrt{\omega}$, in accord with the canonical normalization of the quasidilaton.
Since $\pi$ and $\sigma$ are decoupled in this limit, the solution for $\pi$ is exactly the same as the one in massive gravity, found in Ref. \cite{Berezhiani:2013aa}, while $\sigma$ is just a free massless scalar. The first of the above profiles for $\lambda_1$ corresponds to the self-accelerating background with infinitely strongly coupled vector perturbations, considered in the previous section. We therefore focus on one of the other two solutions\footnote{Considering the other one, $\lambda_1=(1- \sqrt{3})/2\alpha$, will lead to similar conclusions.}, $\lambda_1=(1+ \sqrt{3})/2\alpha$, in which case the next order terms in the $1/\sqrt{w}$ expansion are
\begin{eqnarray}
\lambda_2 &=& \frac{2 \left(\sqrt{3} \alpha ^3-6
	\alpha ^2-3 \alpha +2
	\sqrt{3}+3\right) b}{9 \alpha ^3},\nonumber\\
\tilde\lambda_{\sigma, 2} &=& -\frac{3 \alpha ^3+\left(8+6
	\sqrt{3}\right) \alpha
	^2-\left(17+9 \sqrt{3}\right)
	\alpha +3 \sqrt{3}+5}{6 \alpha ^4}.
\label{lambda2}
\end{eqnarray}
The coefficients of the kinetic and gradient terms in the quadratic perturbation lagrangian for the background of interest are summarized in \ref{sec:perturbation-timedependent}.  As long as $\omega$ is large enough, one can clearly see from these expressions that the conditions for avoiding ghost and gradient instabilities for all helicities are satisfied if
\beq
0<\alpha<\frac{2+\sqrt{3}}{4}~.
\eeq
Moreover, the speed of sound for one combination of the scalar modes and that of the vector helicities, $c_s^{2(-)}$ and $c_{sA}^{2}$, are strictly subluminal, the former propagating at a quarter of the speed of sound. The remaining scalar on the other hand, propagates at the following speed
\begin{eqnarray}
c_s^{2(+)}&=&1+{2 (2+\sqrt{3}-\sqrt{3} \alpha -2\sqrt{3}  \alpha^2 + \alpha^3 )^2\over 3 \alpha ^6 \omega} 
	+{\cal  O}(\omega^{-3/2})~,
\end{eqnarray}
which is always slightly superluminal for large $\omega$.  
Beyond the $1/\omega$ expansion, one can employ numerical analysis to explore  the stable parameter space. The qualitative picture is the same as for the large $\omega$ case: one can readily find a parameter space, devoid of ghosts and gradient instability. Furthermore, all modes propagate at subluminal speed, except for one combination of the scalars, which becomes (exactly luminal) $\sigma$ in the limit of large $\omega$.

The picture remains qualitatively similar for the case of nonzero $\alpha_5$.
To avoid the ghost problem inside the Vainshtein radius, we still need to impose $1+2\alpha a=0$, and then expand the solutions around their $\omega \to \infty$ values, as above. Nonzero $\alpha_5$ does not change the zeroth order background profiles, since it enters only through the potential for $\sigma$, which makes its effects suppressed by powers of $\sqrt{\omega}$. This means that all quantities, possibly except of $c_s^{2(+)}$, determining stability and (sub)luminality of perturbations remain intact, since they are all dominated by the zeroth order contributions. Now, $c_s^{2(+)}$ is exactly one at the zeroth order, and as shown above, receives a slightly superluminal correction in the case of vanishing $\alpha_5$ at $\mathcal{O}(\omega^{-1})$. 
One can straightforwardly convince oneself, that unfortunately the same conclusion persists for $\alpha_5\neq 0$, the speed of sound $c_s^{2(+)}$ being corrected by a positive-definite ($\alpha_5$-dependent) quantity at order $\omega^{-1}$. 
Furthermore, as we checked via numerical analysis, the situation is the same for $\mathcal{O}(1)$, or smaller values of $\omega$, corresponding to stable backgrounds with all modes, but one scalar propagating at superluminal velocity.

\section{Extended theory}
\label{sec6}

We have seen in the previous sections that static solutions, that excite the helicity-zero polarization of the graviton in the original quasidilaton theory are in general problematic, due to the issues with the propagation of ghosts in the Vainshtein region. Mathematically, the problems arise due to the kinetic term of the quasidilaton, ${\cal B}_1=\omega/2$, becoming parametrically suppressed with respect to the mixing with the helicity-0 graviton as $r\ll r_*$. One might therefore think that the situation can be improved by supplementing the $\sigma$ sector by Galileon interactions \cite{Nicolis:2008in} in the decoupling limit, since this would make the kinetic coefficient of the quasidilaton space-dependent, and possibly enhanced within the Vainshtein radius. In addition, we saw that while the ghost problem can be avoided for solutions with decoupled sources (i.e. the ones that \emph{do not} excite the longitudinal graviton), one combination of the scalar modes always propagates at a superluminal speed. This can be seen in the $\omega\to \infty$ limit, by noting that the quasidilaton becomes a free field propagating at exactly the unit speed, while the next order, $1/\omega$ correction to the speed of sound always happens to be in the superluminal direction for the solutions of interest. On the other hand, if the $\omega \to \infty$ limit does not describe a trivial (free) quasidilaton sector, with possibly a nontrivial background $\sigma$-profile characterized by subluminal excitations, one can in principle get rid of superluminal propagation altogether.

Finally, a very important motivation beyond extending the decoupling limit of the original quasidilaton is the realization that the resulting theory can capture all possible technically natural spherically symmetric and static solutions in the most general extensions of the full original quasidilaton (i.e. the theory beyond any limit). Indeed the decoupling limit we will consider in fact represents the most general ghost-free theory of a tensor and a pair of scalars, invariant under internal galilean transformations. This guarantees that any ghost-free modification of the original quasidilaton is bound to reduce to what we consider below at sufficiently short distances. Moreover, the decoupling limit treatment guarantees that the obtained solutions lie well within the regime of validity of the effective theory, and are fully insensitive to any possible UV physics. 

We therefore wish to consider the following action,
\begin{eqnarray}
  &&{\cal L}_{\rm DL}=-{1\over 4} h^{\mu\nu}{\cal E}_{\mu\nu}^{\alpha\beta}h_{\alpha\beta}
  -{1\over 8} \pi\left[\varepsilon\varepsilon \Pi 
    -{2 \alpha \over \Lambda^3}\varepsilon\varepsilon \Pi\Pi
    +{\alpha^2 \over \Lambda^6}
    \varepsilon\varepsilon\Pi\Pi\Pi
  \right]
  \nonumber\\
  &&~~~~~~~~~~~
-\sigma \biggl[
{\omega \over 12}\varepsilon \varepsilon \Sigma
+{\xi_1 \over 6\Lambda^3}\varepsilon \varepsilon \Sigma\Sigma
+{\xi_3 \over 4\Lambda^6}\varepsilon \varepsilon \Sigma\Sigma\Sigma
+{\xi_5 \over 10\Lambda^9}\varepsilon \varepsilon \Sigma\Sigma\Sigma\Sigma
\biggr]
  \nonumber\\
  &&~~~~~~~~~~~
+\sigma\left[4\alpha_5 \Lambda^3+ \gamma_0\varepsilon\varepsilon\Pi
    +{\gamma_1\over \Lambda^3}\varepsilon\varepsilon\Pi\Pi
    +{\gamma_2\over \Lambda^6}\varepsilon\varepsilon\Pi\Pi\Pi
    +{\gamma_3\over \Lambda^9}\varepsilon\varepsilon\Pi\Pi\Pi\Pi\right]
  \nonumber\\
  &&~~~~~~~~~~~
    +{1 \over 2M_{\rm Pl}}h^{\mu\nu}T_{\mu\nu}
    +{1 \over 2M_{\rm Pl}}\pi T
    -{\alpha \over 2M_{\rm Pl}\Lambda^3} \pi \Pi_{\mu\nu}T^{\mu\nu}
\label{LagrangianDLH2}
\end{eqnarray}
which, as shown in appendix \ref{sec:complete-lagrangian-in-the-decoupling-limit}, can be obtained as the decoupling limit of the quasidilaton with ghost-free Horndeski interactions.  Here we have set $\beta=0$ and $\alpha_5=0$ as above, as well as $\xi_2=\xi_4=0$, where $\xi_{2,4}$ are the coefficients in front of the nonlinear interaction terms between the quasidilaton and the helicity-2 graviton in the extended theory.  The parameter choice $\beta=\xi_4=0$ ensures the absence of scalar-tensor interactions unremovable by a tensor mode redefinition, while $\xi_2=0$ removes the disformal coupling of $\sigma$ to the energy momentum tensor. 
The given choice of the model parameters thus corresponds to just the Galileon interactions for $\sigma$, contributed by the additional Horndeski terms. 

The $\pi$-equation of motion, that follows from \eqref{LagrangianDLH2} reads
\begin{eqnarray}
  && -{a \over 2} + 2 b \gamma_0
+\left({3 \over 2} + 3 a \alpha + 4b \gamma_1\right)\lambda
-(3\alpha+3a\alpha^2-6b\gamma_2)\lambda^2
+(\alpha^2+8b\gamma_3)\lambda^3
\nonumber\\
&&  
-(6\gamma_0-4a\gamma_1)\lambda_\sigma
-(8\gamma_1-12a \gamma_2)\lambda_\sigma \lambda
-(6\gamma_2 - 24a \gamma_3)\lambda_\sigma \lambda^2
=(1+2a\alpha)\left({r_* \over r} \right)^3,
  \label{EOM_piH}
\end{eqnarray}
while the equation of motion for $\sigma$ takes on the following form
\begin{eqnarray}
&&4\alpha_5 + 6 a \gamma_0 - b \omega
  -6(3\gamma_0-2a\gamma_1)\lambda
  -6(2\gamma_1-3a\gamma_2)\lambda^2
  -6(\gamma_2-4 a \gamma_3)\lambda^3
\nonumber\\
&&  ~~~~~~~~~~
  +3\omega\lambda_\sigma
  -6b\xi_1 \lambda_\sigma
  +6(\xi_1-3b\xi_3)\lambda_\sigma^2
  +6(\xi_3-2b\xi_5)\lambda_\sigma^3
=0~.  
  \label{EOM_sigmaH}
\end{eqnarray}
In the rest of this section, we will study solutions to these equations analogous to the ones previously obtained, as well as the details of the spectra of perturbations on the corresponding backgrounds. 

\subsection{Time-independent solutions}
\label{sec6.1}

We start out by considering time-independent Vainshtein solutions. The simplest extension of our previous analysis would correspond to setting $\xi_3=\xi_5=0$, in which case the solution inside the Vainshtein radius gives, $\lambda \propto (r_*/r)^{6/7}$ and $\lambda_\sigma \propto (r_*/r)^{9/7}$.  This yields the following $r\ll r_*$ behavior of the kinetic coefficients in the quadratic perturbation action for the scalar modes, ${\cal B}_1 \propto (r_*/r)^{9/7}$, ${~\cal C}_1^2/{\cal A}_1 \propto (r_*/r)^3$, leading again to a scalar ghost in the Vainshtein region. This, as we now show, can be avoided upon inclusion of the quartic Galileon, $\xi_3 \neq 0$. 
To make things simple and analytic, we again consider the further limit, in which $\pi$ and $\sigma$ are decoupled. To this end, one can again set $\omega \to \infty$ just as we did in the previous section; however, in contrast to the previous case, we'd like the resulting $\sigma$ sector to retain Galileon interactions in order to allow for non-trivial backgrounds. This requires to scale the ${\tilde \xi}$ coefficients accordingly, the proper limit defined as follows
\begin{eqnarray}
\omega \to \infty,\qquad 
{\tilde \xi}_1 \equiv {\xi_1 \over \omega^{3/2}}={\rm finite},\qquad 
{\tilde \xi}_3 \equiv {\xi_3 \over \omega^{2}}={\rm finite},\qquad 
{\tilde \xi}_5 \equiv {\xi_5 \over \omega^{5/2}}={\rm finite}~.
\end{eqnarray}
The action (\ref{LagrangianDLH2}) then splits up in this limit into separate, non-interacting Galileon theories for $\pi$ and $\sigma$. For simplicity, we further impose the condition that the two Galileon sectors are of similar structure. This can be achieved by requiring
\begin{eqnarray}
\xi_3={\xi_1^2 \over 3\omega}, \qquad \xi_5=0~,
\label{symmetricCondition}
\end{eqnarray}
which makes the two sectors symmetric under the interchange $(\pi \leftrightarrow \tilde\sigma, ~\alpha \leftrightarrow -{\tilde \xi_1})$, apart form the source term in the equation of motion for $\pi$. 
We then expand the solutions in terms of $\omega^{-1/2}$ as follows
\begin{eqnarray}
&&\lambda = x_1 + x_2 \omega^{-1/2}+{\cal O}(\omega^{-1}), 
\qquad {\tilde \lambda}_\sigma = y_1 + y_2 \omega^{-1/2} +{\cal O}(\omega^{-1}),
\end{eqnarray}
where $x_1$ and $y_1$ are determined from,
\begin{eqnarray}
2\alpha^2 x_1^3 -6 \alpha x_1^2 +3 x_1&=& 2\({r_* \over r}\)^2~,\\
2 \xi_1^2 y_1^3+6\xi_1 y_1^2 + 3 y_1&=&0~,
\end{eqnarray}
The expressions for $x_2$ and $y_2$ can then be obtained perturbatively.
In general, there are multiple solutions within the Vainshtein radius, out of which we will focus on the following one
\begin{eqnarray}
x_1&=& {1 \over \alpha^{2/3}} {r_* \over r} +{1 \over \alpha} +{1 \over 2 \alpha^{4/3}} {r \over r_*}  + {\cal O}\({r \over r_*}\)^2\\
y_1&=& - {3 +\sqrt{3} \over 2 {\bar \xi_1}}\\
x_2&=& - {(3 +\sqrt{3}) \gamma_2 \over \alpha^2 {\bar \xi_1}}  - {2(3 +\sqrt{3}) (2\alpha \gamma_1 + 3\gamma_2) \over 3\alpha^{7/3} {\bar \xi_1}} {r \over r_*} + {\cal O}\({r \over r_*}\)^2
\\
y_2&=& 	{2 \gamma_2 \over (1+\sqrt{3}) \alpha^2} \({r_* \over r} \)^3+	{2 (2 \alpha\gamma_1+3\gamma_2) \over (1+\sqrt{3})\alpha^{7/3} } \({r_* \over r} \)^2 + {\cal O}\({r_* \over r}\)~.\\
\end{eqnarray}
The leading terms in the expression for $x_1$  and $y_1$ correspond to the `restricted Galileon' discussed in \cite{Berezhiani:2013aa}.
Since there is no source term in the $\sigma$-equation, the $\omega \to \infty$ solution describes a cosmological background, $\lambda_\sigma={\rm const}$; couplings between the $\pi$ and $\sigma$ sectors on the other hand introduce weak space dependence in $\lambda_\sigma$ within the Vainshtein radius.
Once substituted into the expressions for the kinetic coefficients of the quadratic perturbation action, the above background solution yields
\begin{eqnarray}
{\cal A}_1 &=&\[{3 \alpha^{2/3} \over 2}\({r_* \over r}\)^2 + {\cal O}\({r_* \over r}\)\] \omega^0 + {\cal O}(\omega^{-1/2}),\\ 
{\cal B}_1 - \frac{{\cal C}_1^2}{4{\cal A}_1} &=& 5+3\sqrt{3} + {\cal O}(\omega^{-1/2})~.
\end{eqnarray}
For positive $\alpha$, the solution at hand is free from scalar ghosts.
The radial and angular sound speeds can be evaluated following the procedure, outlined in Appendix \ref{sec:scalar-perturbations}
\begin{eqnarray}
c_r^{2(+)}&=&\[ 1- {2 \over \alpha^{1/3}} {r \over r_*}+ {\cal O}\({r \over r_*}\)^2~\] \omega^0
+{\cal O}(\omega^{-1/2}),\label{cr2+}\\
c_r^{2(-)}&=&  1-{\sqrt{3} \over 2} +{\cal O}(\omega^{-1/2}), \label{cr2-}\\
c_\Omega^{2(+)}&=& \[ {1 \over \alpha^{2/3}} \({r \over r_*}\)^2+ {\cal O}\({r \over r_*}\)^3~ \] \omega^0
+{\cal O}(\omega^{-1/2}),\label{cO2+}\\
c_\Omega^{2(-)}&=& 1-{\sqrt{3} \over 2} +{\cal O}(\omega^{-1/2})\label{cO2-}.
\label{angularSS}
\end{eqnarray}
All these expressions are manifestly positive for large $\omega$. Furthermore, the radial sound speed for $\pi$ (\ref{cr2+}) is slightly subluminal while the angular speed (\ref{cO2+}) is suppressed by the small factor $(r / r_*)^2$, making it extremely subluminal inside the Vainshtein radius (this is generic to Vainshtein solutions in Galileon theories \cite{Nicolis:2008in}). Both radial and angular sound speed for $\sigma$, (\ref{cr2-}) and (\ref{cO2-}), are also subluminal, $c_{r,~\Omega}^{~2(-)} \approx 0.134$.
The kinetic coefficients and the sound speeds in various directions of the vector perturbations are given as follows
\begin{eqnarray}
C_{tr} &=& \[{\alpha^{4/3} \over (1-2\alpha)} {r_*\over r}+ {\cal O}(r^0)\]\omega^0+ {\cal O}(\omega^{-1/2}),\\ 
C_{t\theta} &=&\[ {\alpha \over 2}+ {\cal O}(r)\]\omega^0+ {\cal O}(\omega^{-1/2}),\\
c_r^{2(A)}&=& \[1-{2\over \alpha^{1/3}} {r\over r_*}+ {\cal O}(r^2)\]\omega^0+ {\cal O}(\omega^{-1/2}),\\
c_{\Omega,1}^{2(A)}&=&\[{1-2\alpha \over 2 \alpha^{1/3}} {r\over r_*}+ {\cal O}(r^2)\]\omega^0+ {\cal O}(\omega^{-1/2}), \\
c_{\Omega,2}^{2(A)}&=&\[{1 \over 2 \alpha^{2/3}} \( {r\over r_*}\)^2+ {\cal O}(r^3)\]\omega^0+ {\cal O}(\omega^{-1/2}).
\end{eqnarray}
The latter expressions show that the vector perturbations are also free of ghosts, gradient instabilities, and superluminal propagation for large $\omega$, and as long as $0 < \alpha<1/2$ is satisfied. 

We proceed by looking at the behaviour of the solutions at large distances.  The asymptotically Minkowski solution is the same as the one in (\ref{Minkowsol}), which as we have argued above, leads to ghosts inside reasonable sources. We thus disregard this branch, moving to the asymptotically curved, cosmological solutions in the $\omega \to \infty$ limit.  The field profiles in this limit read 
\begin{eqnarray}
\lambda_{\rm decoupled}=0, ~\frac{3 \pm \sqrt{3}}{2\alpha}, \qquad
{\tilde \lambda}_{\sigma,{\rm decoupled}}= 0, ~- \frac{3 \pm \sqrt{3}}{2 {\tilde \xi_1}}~,
\end{eqnarray}
and they are identical due to the $\alpha\to -\tilde \xi_1$ interchange symmetry we have imposed above. 
The solution $\lambda_{\rm decoupled}=(3 - \sqrt{3}) /2\alpha$ leads to the wrong sign for the kinetic term of $\pi$-perturbations, ${\cal A}_1 = 3(5-3\sqrt{3})/2<0$, we therefore disregard it (and its dual in the $\sigma$ sector), and concentrate on the only remaining solution for $\lambda_{\rm decoupled}$ (this is also the one that matches the field profiles inside the Vainshtein region, obtained above). Furthermore, we will concentrate on the only remaining nozero solution for the quasidilaton, corresponding to ${\tilde \lambda}_{\sigma,{\rm decoupled}}=-(3 + \sqrt{3}) /2\tilde \xi_1$. 

The perturbations over the obtained background can be treated along the lines of what we did in the previous section in eq. (\ref{expandedsigma}). The expressions for quantities, determining stability and the speed of propagation of fluctuations of various helicity read
\begin{eqnarray}
{\cal A}_1 &=&  \frac{3}{2} \left(5+3 \sqrt{3}\right)+ {\cal O}(\omega^{-1/2}),\\
{\cal B}_1-{{\cal C}_1^2 \over 4{\cal A}_1}  &=& 5+3 \sqrt{3}+ {\cal O}(\omega^{-1/2}),\\
c_s^{2(+)}&=&1-\frac{\sqrt{3}}{2}+ {\cal O}(\omega^{-1/2}),\\
c_s^{2(-)}&=&1-\frac{\sqrt{3}}{2}+ {\cal O}(\omega^{-1/2}),\\
{\bar D}_1&=&  \frac{\left(2+\sqrt{3}\right)	\alpha }{3+\sqrt{3}-4 \alpha} + {\cal O}(\omega^{-1/2}),\\
c_s^{2(A)}&=& \frac{-2 \sqrt{3} \alpha +2 \alpha +\sqrt{3}}{6+2 \sqrt{3}-4 \alpha }+ {\cal O}(\omega^{-1/2})~.
\end{eqnarray}
One can see, that for $0<\alpha < (3+\sqrt{3})/4$ and for sufficiently large $\omega$, the system is free from any sort of instability and all speeds of sound are safely subluminal.

\subsection{Time-dependent solutions}

We have seen that extending the quasidilaton by Horndeski-like terms can take care of stability problems associated with time-independent solutions in the original theory. 
In principle, for a certain subclass of time-dependent solutions one can again use the same arguments as before:  the stable solutions (both inside and outside the Vainshtein radius) can be explicitly constructed by expanding around the time-independent solutions obtained in Sec.~\ref{sec6.1} as long as the contributions from the time-dependent pieces of the scalar fields, $a$ and $b$, are small. Therefore, the time-dependent solutions are also free of ghosts, tachyons, gradient instabilities, and superluminal modes in the extended quasidilaton theory.

\begin{figure}[t]
	\begin{center}
		\includegraphics[scale=1.2]{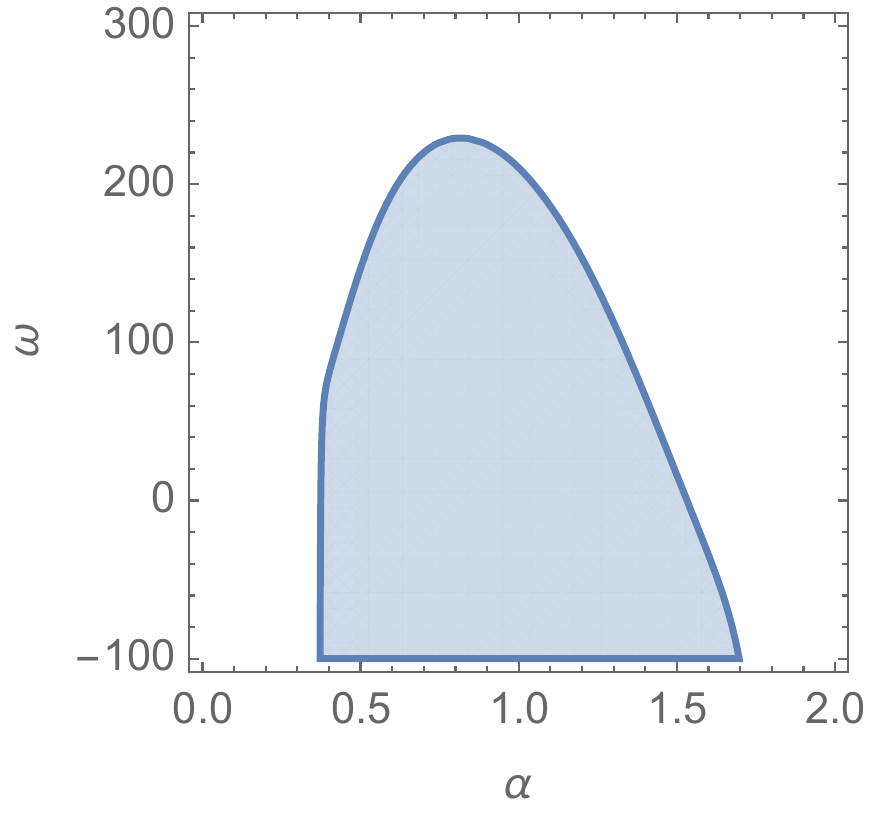}
	\end{center}
	\caption{This plot shows 
		the condition for avoiding ghosts, tachyon, gradient instabilities, superluminal modes for the time-dependent solution with $1+\alpha a=0$.
		The parameters are chosen to be $\lambda=3$, $\lambda_\sigma=1$, and $\xi_3=50$.
	}
	\label{fig:TimedependentHorn}
\end{figure}

Next we would like to focus the special case with decoupled sources, analogous to the one of Sec. \ref{sec5}.
To this end, we again focus on solutions corresponding to $1+\alpha a=0$. Moreover, to simplify the analysis we will again impose symmetry under $\pi \leftrightarrow \tilde\sigma$, which, in addition to (\ref{symmetricCondition}), requires $b={{\omega} / 2\xi_1}$ to hold. The relevant solution in the $\omega \to \infty$ limit then reads \cite{Berezhiani:2013aa}
\begin{eqnarray}
\label{solstd}
  \lambda_{\rm decoupled}={1\over 2\alpha}, ~\frac{1 \pm \sqrt{3}}{2\alpha}, \qquad
  {\tilde \lambda}_{\sigma,{\rm decoupled}}=- {1\over 2\xi_1}, ~- \frac{1 \pm \sqrt{3}}{2 {\tilde \xi_1}}
\end{eqnarray}
The solution $\lambda_{\rm decoupled}={1/ 2\alpha}$ corresponds to a de Sitter background with infinitely strongly coupled vector perturbations, as shown above. We therefore disregard it. 
The analysis of perturbations in $1/\omega$ expansion for one representative background from \eqref{solstd}, corresponding to $\lambda_{\rm decoupled}=(1 + \sqrt{3})/(2\alpha),~{\tilde \lambda}_{\sigma,{\rm decoupled}}= - (1 - \sqrt{3})(2 {\tilde \xi_1})$ , is given in Appendix \ref{sec:perturbation-timedependent-Horn}. Most importantly, the $\omega\to\infty$ solutions are free of ghosts, gradient instability and superluminal propagation for all modes present in the theory for
\beq
0<\alpha < \frac{1}{4} (2+\sqrt{3})~.
\eeq
Since perturbations are stable and safely subluminal for $\omega\to \infty$, we expect that they generically remain such at least down to moderate values of $\omega$ (when all other parameters are taken of order unity).

Finally, for illustrative purposes we provide a different, numerically obtained example of a parameter space completely free of all kinds of instability and superluminal propagation. 
Rather than solving for the quantities $\lambda$ and $\lambda_\sigma$, we treat these as input parameters, and solve for $\xi_1$ and $b$
instead (we have assumed $\alpha_5=\xi_5=0$ in this analysis). The results, displayed on Fig. \ref{fig:TimedependentHorn} confirm that there is typically a rich parameter space for stable and subluminal backgrounds.

%

\section{Conclusions}
In this paper, we have investigated spherically symmetric solutions in the decoupling limit 
of quasidilaton theory in the absence of the un-diagonalizable interaction term, 
$h^{\mu\nu}\varepsilon_\mu\varepsilon_\nu \Pi\Pi\Pi$.
In the decoupling limit of the \emph{original} quasi-dilaton theory (\ref{LagrangianDL2}), we have found that 
\begin{itemize}
	\item Both of the scalar perturbations become ghosts inside the Vainshtein radius for 
	spherically symmetric configurations that asymptote to Minkowski space at infinity. 
	\item The only option to avoid ghosts is to consider solutions that correspond to $1+\alpha a =0$, 
	which leads to vanishing of the source term in the $\pi$  equation of motion, and to a modification of the boundary conditions at infinity.  
	In this case, there are no  ghosts, tachyons or gradient instability, but  one scalar 
	mode always propagates with a superluminal group velocity.
\end{itemize}
Thus the condition $1+\alpha a =0$ is crucial in order to have stable solutions in the original quasidilaton theory, where 
any asymptotically-flat background always suffers from ghost excitations in the 
Vainshtein region.

Furthermore,  we show that a general self-accelerated solution found in  \cite{Gabadadze:2014aa}, when 
restricted to a special case of $\beta=0$,  shares the same problem: the kinetic term of 
the scalar field, which has a right sign on a de Sitter background, flips the sign inside the Vainshtein radius.
This means that the sign of the kinetic term vanishes somewhere as one approaches a source from far away; 
at that point one of the scalar fields  becomes infinitely strongly coupled and the classical solution -- meaningless. 

A way to avoid ghost instability inside the Vainshtein solution is to supplement the theory by shift-symmetric Horndeski 
terms for the $\sigma$ field;  the latter are  naturally allowed by the quasidilaton symmetry (\ref{Symmetry}). By adding these terms, the $\sigma$ field acquires the cubic, quartic, and quintic Galileon self-interactions in the decoupling limit, which can in principle cure the ghost problem inside the Vainshtein radius.  In particular,  we find  in this case, that 
\begin{itemize}
	\item The solution approaching  Minkowski spacetime at large distances cannot be allowed due to the presence of a ghost in the scalar sector.
	
	\item There exists another branch of solutions with cosmological   
	asymptotics at large distances, that is   free of ghosts, tachyons, gradient instability,  and 
	superluminal propagation. The  latter class of solutions are of a highest interest.
\end{itemize}

As we pointed out above, throughout this paper we  set the coefficient of the 
interaction term, $h^{\mu\nu}\varepsilon_\mu\varepsilon_\nu \Pi\Pi\Pi$, to zero. 
This is a technically natural choice \cite{Rham:2012aa}, 
however, inclusion of this term may open novel 
branches of solutions and should be investigated in future.



\acknowledgments 
GG's work is  supported by NASA grant NNX12AF86G S06, and NSF grant PHY-1316452. 
R.K. is supported in part by JSPS Postdoctoral Fellowships for Research Abroad. The work of D.P. is supported in part by MIUR-FIRB grant RBFR12H1MW and by funds provided by Scuola Normale Superiore through the program "Progetti di Ricerca per Giovani Ricercatori".


\appendix

\section{Complete Lagrangian in the decoupling limit}\label{sec:complete-lagrangian-in-the-decoupling-limit}
In this appendix we consider the action of quasi-dilaton theory 
supplemented by the shift-symmetric Horndeski Lagrangian, given as follows
\begin{eqnarray}
S_{\rm H}= \int d^4x \sqrt{-g} \, 
\sum_{i=2}^5 {\cal L}_i,
\label{ActionHorndeski}
\end{eqnarray}
where 
\begin{eqnarray}
&&{\cal L}_2=P(X),\nonumber\\
&&{\cal L}_3=-G_3(X)\square \sigma\nonumber,\\
&&{\cal L}_4=G_4(X) R 
+ G_{4X}\left[ (\square \sigma)^2-(\nabla_\mu\nabla_\nu \sigma)^2\right]\nonumber,\\
&&{\cal L}_5=G_5(X) G_{\mu\nu} \nabla^\mu\nabla^\nu \sigma
-{1 \over 6} G_{5X}(X) \biggl[ (\square \sigma)^3
-3\square \sigma(\nabla_\mu\nabla_\nu \sigma)^2+2(\nabla_\mu\nabla_\nu \sigma)^3\biggl].
\label{ActionHorndeski}
\end{eqnarray}
Here $P(X)$ and $G_i(X)$ are arbitrary functions of $X=-(\partial \sigma)^2/2$,
and $G_{iX}$ denotes the derivative of $G_i$ with respect to $X$, 
$\partial G_{i}/ \partial X$.
This action is invariant under constant shifts (\ref{Symmetry}).

As a next step, we derive the complete scalar-tensor Lagrangian in the decoupling limit
of the quasidilaton theory including the shift-symmetric Horndeski terms.
The decoupling limit of the Horndeski theory has been  
derived in \cite{Koyama:2013aa}, and we follow their approach.
To find the effective action in the decoupling limit, 
we first perturb the quasi-dilaton around a constant background value, $\sigma \to \sigma_0 + \sigma$ 
as well as perturb the metric around Minkowski spacetime,
$g_{\mu\nu}=\eta_{\mu\nu}+h_{\mu\nu}$.
We then Taylor-expand the entire action and keep the
dominant interactions for the Vainshtein mechanism,
schematically given as 
$\sigma (\partial \sigma)^n$ and $h (\partial \sigma)^2$.
Here we assume that the terms such as 
$(\partial \sigma)^3$, $(\partial \sigma)^4$, etc., are
suppressed with respect to galileon;like interactions.
The complete effective Lagrangian is then given by
\begin{eqnarray}
  &&{\cal L}_{\rm DL}^{( h,\pi,\sigma)}
=-{1\over 4} h^{\mu\nu}{\cal E}_{\mu\nu}^{\alpha\beta}h_{\alpha\beta}
  -h^{\mu\nu} \left[{1\over 4}\varepsilon_\mu\varepsilon_\nu \Pi 
    -{\alpha \over 4 \Lambda^3}\varepsilon_\mu\varepsilon_\nu \Pi\Pi
    -{\beta\over 2\Lambda^6}\varepsilon_\mu\varepsilon_\nu \Pi\Pi\Pi\right]\nonumber\\
  &&~~~~~~~~~~~
+\sigma\left[4\alpha_5 \Lambda^3+ \gamma_0\varepsilon\varepsilon\Pi
    +{\gamma_1\over \Lambda^3}\varepsilon\varepsilon\Pi\Pi
    +{\gamma_2\over \Lambda^6}\varepsilon\varepsilon\Pi\Pi\Pi
    +{\gamma_3\over \Lambda^9}\varepsilon\varepsilon\Pi\Pi\Pi\Pi\right]\nonumber\\
  &&~~~~~~~~~~~
-\sigma \biggl[
{\omega \over 12}\varepsilon \varepsilon \Sigma
+{\xi_1 \over 6\Lambda^3}\varepsilon \varepsilon \Sigma\Sigma
+{\xi_3 \over 4\Lambda^6}\varepsilon \varepsilon \Sigma\Sigma\Sigma
+{\xi_5 \over 10\Lambda^9}\varepsilon \varepsilon \Sigma\Sigma\Sigma\Sigma
\biggr]
\nonumber\\
  &&~~~~~~~~~~~
  +h^{\mu\nu} \left[
    {\xi_2 \over 2 \Lambda^3}\varepsilon_\mu\varepsilon_\nu \Sigma\Sigma
    -{\xi_4 \over 2\Lambda^6}\varepsilon_\mu\varepsilon_\nu \Sigma\Sigma\Sigma\right]\nonumber\\
\nonumber\\
  &&~~~~~~~~~~~
  +{1 \over 2M_{\rm Pl}}h^{\mu\nu}T_{\mu\nu},
\label{LagrangianDLH}
\end{eqnarray}
where we have introduced the dimensionless parameters $\xi_i$ as follows
\begin{eqnarray}
  &&\omega=P_X,\qquad
  \xi_1=-G_{3X} \Lambda^3,\qquad
  \xi_2={\Lambda^3 \over \mpl} G_{4X},\qquad\nonumber \\
  &&\xi_3=G_{4XX} \Lambda^6,\qquad
  \xi_4=-{\Lambda^6 \over 3\mpl} G_{5X},\qquad
  \xi_5=-{1\over 3} G_{5XX} \Lambda^9,\qquad
\end{eqnarray}
these being evaluated at the background value,
i.e., $X=0$.
The Lagrangian is still invariant under the gauge transformation, $h_{\mu\nu} \to h_{\mu\nu} + \partial_{\mu}\xi_{\nu}+\partial_{\nu}\xi_{\mu}$ and the galileon transformation for $\pi$ and $\sigma$ fields, $\partial_\mu \pi \to \partial_\mu \pi + c_\mu$ and $\partial_\mu \sigma \to \partial_\mu \sigma + d_\mu$.
Here the "k-essence" term, $P(X)$, gives 
the standard kinetic term $(\partial \sigma)^2$ in the decoupling limit;
this defines $\omega$ in (\ref{Action}).
Note that the lowest possible interaction between $h$ and $\sigma$, $h_{\mu\nu}\varepsilon_\mu\varepsilon_\nu\Sigma$, is absent unlike 
for the $\pi$ field. This interaction could be obtained from
e.g. a term like $\sigma R$; the shift-symmetry
in the $\sigma$ sector however forbids such an interaction.

The Horndeski terms introduce the $h$-$\sigma$ couplings, similar to the 
$h$-$\pi$ terms of pure massive gravity.
 In the similar way, 
these couplings can be removed by 
a local field redefinition,
\begin{eqnarray}
  h_{\mu\nu} \to h_{\mu\nu}+\pi\eta_{\mu\nu}
  -{\alpha \over \Lambda^3} \pi \Pi_{\mu\nu}
  -{2\xi_2 \over \Lambda^3} \sigma \Sigma_{\mu\nu},
\end{eqnarray}
except for the $h^{\mu\nu}\varepsilon_\mu\varepsilon_\nu \Pi\Pi\Pi$ and 
$h^{\mu\nu}\varepsilon_\mu\varepsilon_\nu \Sigma\Sigma\Sigma$ terms.
Then the action in the new frame becomes
\begin{eqnarray}
  &&{\cal L}_{\rm DL}=-{1\over 4} h^{\mu\nu}{\cal E}_{\mu\nu}^{\alpha\beta}h_{\alpha\beta}
  \nonumber\\
  &&~~~~~~~~~~~
  -{1\over 8} \pi\left[\varepsilon\varepsilon \Pi 
    -{2 \alpha \over \Lambda^3}\varepsilon\varepsilon \Pi\Pi
    +{\alpha^2 - 4 \beta \over \Lambda^6}
    \varepsilon\varepsilon\Pi\Pi\Pi
    +{4\alpha\beta \over \Lambda^9}\varepsilon\varepsilon \Pi\Pi\Pi\Pi
  \right]
  \nonumber\\
  &&~~~~~~~~~~~
-\sigma \biggl[
{\omega \over 12}\varepsilon \varepsilon \Sigma
+{\xi_1 \over 6\Lambda^3}\varepsilon \varepsilon \Sigma\Sigma
+{2\xi_2^2+\xi_3 \over 4\Lambda^6}\varepsilon \varepsilon \Sigma\Sigma\Sigma
+{10\xi_2\xi_4+\xi_5 \over 10\Lambda^9}\varepsilon \varepsilon \Sigma\Sigma\Sigma\Sigma
\biggr]
  \nonumber\\
  &&~~~~~~~~~~~
+\sigma\left[4\alpha_5 \Lambda^3+ \gamma_0\varepsilon\varepsilon\Pi
    +{\gamma_1\over \Lambda^3}\varepsilon\varepsilon\Pi\Pi
    +{\gamma_2\over \Lambda^6}\varepsilon\varepsilon\Pi\Pi\Pi
    +{\gamma_3\over \Lambda^9}\varepsilon\varepsilon\Pi\Pi\Pi\Pi\right]
  \nonumber\\
  &&~~~~~~~~~~~
+\pi\left[
    {\xi_2 \over 2\Lambda^3}\varepsilon\varepsilon\Sigma\Sigma
    +{\xi_4\over 2\Lambda^6}\varepsilon\varepsilon\Sigma\Sigma\Sigma
    -{\alpha \xi_2\over 2\Lambda^6}\varepsilon\varepsilon\Pi\Sigma\Sigma
    -{\alpha \xi_4 \over 2\Lambda^9}\varepsilon\varepsilon\Pi\Sigma\Sigma\Sigma
    -{\beta \xi_2 \over \Lambda^9}\varepsilon\varepsilon\Pi\Pi\Sigma\Sigma
\right]
  \nonumber\\
  &&~~~~~~~~~~~
  +{\beta\over 2\Lambda^6}h^{\mu\nu}\varepsilon_\mu\varepsilon_\nu \Pi\Pi\Pi
  +{\xi_4 \over 2\Lambda^6}h^{\mu\nu}\varepsilon_\mu\varepsilon_\nu \Sigma\Sigma\Sigma
  \nonumber\\
  &&~~~~~~~~~~~
    +{1 \over 2M_{\rm Pl}}h^{\mu\nu}T_{\mu\nu}
    +{1 \over 2M_{\rm Pl}}\pi T
    -{\alpha \over 2M_{\rm Pl}\Lambda^3} \pi \Pi_{\mu\nu}T^{\mu\nu}
    -{\xi_2 \over M_{\rm Pl}\Lambda^3} \sigma \Sigma_{\mu\nu}T^{\mu\nu}.
\end{eqnarray}
The equation of motion for $h_{\mu\nu}$ reads
\begin{eqnarray}
 {\cal E}_{\mu\nu}^{\alpha\beta}h_{\alpha\beta}={\beta \over \Lambda^6} \varepsilon_\mu\varepsilon_\nu \Pi\Pi\Pi + {\xi_4 \over \Lambda^6 } \varepsilon_\mu\varepsilon_\nu \Sigma\Sigma\Sigma
+{1 \over M_{\rm Pl}}T_{\mu\nu},
\end{eqnarray}
while the equations of motion for $\pi$ and $\sigma$ yield respectively
\begin{eqnarray}
&&{1\over 4}\varepsilon\varepsilon \Pi 
-{3\alpha \over 4 \Lambda^3}\varepsilon\varepsilon \Pi\Pi
+{\alpha^2-4\beta \over 2\Lambda^6}\varepsilon\varepsilon \Pi\Pi\Pi
+{5\alpha \beta \over 2\Lambda^9}\varepsilon\varepsilon \Pi\Pi\Pi\Pi
\nonumber\\
&&~~~~~~~~~
-\gamma_0\varepsilon\varepsilon\Sigma
-{2\gamma_1\over \Lambda^3}\varepsilon\varepsilon\Sigma\Pi
-{3\gamma_2\over \Lambda^6}\varepsilon\varepsilon\Sigma\Pi\Pi
-{4\gamma_3\over \Lambda^9}\varepsilon\varepsilon\Sigma\Pi\Pi\Pi
\nonumber\\
&&~~~~~~~~~
-{\xi_2 \over 2 \Lambda^3}\varepsilon\varepsilon \Sigma\Sigma
-{\xi_4 \over 2\Lambda^6}\varepsilon\varepsilon \Sigma\Sigma\Sigma
+{\alpha \xi_2 \over \Lambda^6}\varepsilon\varepsilon \Pi\Sigma\Sigma
+{\alpha\xi_4 \over \Lambda^9}\varepsilon\varepsilon\Pi\Sigma\Sigma\Sigma
+{3\beta\xi_2 \over \Lambda^9}\varepsilon\varepsilon\Pi\Pi\Sigma\Sigma
\nonumber\\
&&~~~~~~~~~
-{3 \beta \over 2\Lambda^6}\varepsilon^{\mu\alpha\rho\gamma}\varepsilon^{\nu\beta\sigma\gamma} \partial_\mu\partial_\nu h_{\mu\nu} \Pi_{\rho\sigma}\Pi_{\gamma\delta}
={1 \over 2M_{\rm Pl}}T-{\alpha \over M_{\rm Pl}\Lambda^3}\Pi_{\mu\nu}T^{\mu\nu},
\end{eqnarray}
\begin{eqnarray}
&&-{\omega \over 6} \varepsilon\varepsilon\Sigma
-{\xi_1 \over 2\Lambda^3}\varepsilon \varepsilon \Sigma\Sigma
-{2\xi_2^2 + \xi_3 \over \Lambda^6}\varepsilon \varepsilon \Sigma\Sigma\Sigma
-{10\xi_2\xi_4+ \xi_5 \over 2\Lambda^9}\varepsilon \varepsilon \Sigma\Sigma\Sigma\Sigma
\nonumber\\
&&~~~~~~~~~
+4\alpha_5 \Lambda^3+\gamma_0\varepsilon\varepsilon\Pi
    +{\gamma_1\over \Lambda^3}\varepsilon\varepsilon\Pi\Pi
    +{\gamma_2\over \Lambda^6}\varepsilon\varepsilon\Pi\Pi\Pi
    +{\gamma_3\over \Lambda^9}\varepsilon\varepsilon\Pi\Pi\Pi\Pi
\nonumber\\
&&~~~~~~~~~
+{\xi_2 \over  \Lambda^3}\varepsilon\varepsilon \Pi\Sigma
+{3\xi_4 \over 2\Lambda^6}\varepsilon\varepsilon \Pi\Sigma\Sigma
-{\alpha \xi_2 \over 2\Lambda^6}\varepsilon\varepsilon \Pi\Pi\Sigma
-{\alpha\xi_4 \over 2\Lambda^9}\varepsilon\varepsilon\Pi\Pi\Sigma\Sigma
-{2\beta\xi_2 \over \Lambda^9}\varepsilon\varepsilon\Pi\Pi\Pi\Sigma
\nonumber\\
&&~~~~~~~~~
+{3 \xi_4 \over 2\Lambda^6}\varepsilon^{\mu\alpha\rho\gamma}\varepsilon^{\nu\beta\sigma\gamma} \partial_\mu\partial_\nu h_{\mu\nu} \Sigma_{\rho\sigma}\Sigma_{\gamma\delta}
={2\xi_2 \over M_{\rm Pl}\Lambda^3}\Sigma_{\mu\nu}T^{\mu\nu}.
\end{eqnarray}

\section{Scalar Perturbations}\label{sec:scalar-perturbations}
In this appendix, we summarize the quadratic Lagrangian of the fluctuations of
$\pi$ and $\sigma$ around the background (\ref{backgroundPiSigma}).
We define the fluctuations $\phi$ and $\psi$ as
\begin{eqnarray}
  &&\pi(t,x)  \to {a \over 2} \Lambda^3 t^2 + \pi(r) + \phi(t,x)\nonumber\\
  &&\sigma(t,x) \to {b \over 2} \Lambda^3 t^2 + \sigma(r) +\psi(t,x).
\label{perturbations}
\end{eqnarray}
Then the quadratic Lagrangian becomes
\begin{eqnarray}
  {\cal L}^{(2)}_{\rm DL}&=&
  {\cal A}_1 (\partial_t \phi)^2 -{\cal A}_2 (\partial_r \phi)^2
  -{\cal A}_3 (\partial_\Omega \phi)^2 
  +  {\cal B}_1 (\partial_t \psi)^2 -{\cal B}_2 (\partial_r \psi)^2
  -{\cal B}_3 (\partial_\Omega \psi)^2 \nonumber\\
  &&+  {\cal C}_1 \partial_t \phi \, \partial_t \psi
  -{\cal C}_2 \partial_r \phi \, \partial_r \psi
  -{\cal C}_3 \partial_\Omega \phi \, \partial_\Omega \psi,
\label{ActionQuadratic}
\end{eqnarray}
where the coefficients are 
\begin{eqnarray}
  &&{\cal A}_1 ={3 \over 4} 
  -{1 \over \Lambda^3}\left[{3\over 2} \alpha \left(\pi''+2{\pi' \over r}\right)
  +2\gamma_1\left(\sigma''+2{\sigma' \over r}\right)\right]\nonumber\\
&&~~~~~~~~
  +{1 \over \Lambda^6}\left[
  {3\over 2}\alpha^2
  \left({\pi'^2 \over r^2}+2{\pi'\pi'' \over r}\right)
  -6\gamma_2\left({\sigma'\pi' \over r^2}+{\sigma''\pi' \over r}
    +{\sigma'\pi'' \over r}\right)\right]\nonumber\\
&&~~~~~~~~
  -{12\gamma_3 \over \Lambda^9}\left( {\sigma''\pi'^2 \over r^2}+2{\sigma'\pi'\pi'' \over r^2}\right),\\
  &&{\cal A}_2 ={3 \over 4}+ {3 \over 2}\alpha a+2\gamma_1 b 
  -{1 \over \Lambda^3}\left[3(\alpha+\alpha^2 a -2\gamma_2 b) {\pi' \over r}
  +2(2\gamma_1-3\gamma_2 a){\sigma' \over r}\right]\nonumber\\
  &&~~~~~~~~
  +{1 \over \Lambda^6}\left[
  {3\over 2}\left(\alpha^2+8\gamma_3 b\right){\pi'^2 \over r^2}
  -6(\gamma_2-4\gamma_3 a){\sigma'\pi' \over r^2}\right],\\
  &&{\cal A}_3 ={3 \over 4} + {3\over 2} \alpha a + 2\gamma_1 b\\
  &&~~~~~~~~
  -{1 \over \Lambda^3}\left[{3\over 2} (\alpha+\alpha^2 a - 2\gamma_2 b) 
    \left(\pi''+{\pi' \over r}\right)
  +(2\gamma_1-3\gamma_2 a)\left(\sigma''+{\sigma' \over r}\right)\right]  \nonumber\\
  &&~~~~~~~~
+{1 \over \Lambda^6}\left[
  {3\over 2}\left(\alpha^2+8\gamma_3 b\right){\pi'\pi'' \over r}
  -3(\gamma_2-4\gamma_3 a) {\pi'\sigma''+\sigma'\pi'' \over r}\right],\\
  &&{\cal B}_1 ={\omega \over 2}
  +{\xi_1 \over \Lambda^3} \left(\sigma''+2{\sigma' \over r}\right)
  +{3\xi_3 \over \Lambda^6} \left({\sigma'^2 \over r^2}+2{\sigma'\sigma'' \over r}\right)
  +{6\xi_5 \over \Lambda^9}{\sigma'^2 \sigma'' \over r^2},\\
  &&{\cal B}_2 ={\omega \over 2} -\xi_1 b
  +{2 (\xi_1 -3 \xi_3 b )\over \Lambda^3} {\sigma' \over r}
  +{3(\xi_3-2\xi_5 b) \over \Lambda^6} {\sigma'^2 \over r^2},\\
  &&{\cal B}_3 ={\omega \over 2} -\xi_1 b
  +{\xi_1-3\xi_3 b \over \Lambda^3}  \left(\sigma''+{\sigma' \over r}\right)
  +{3(\xi_3-2\xi_5 b) \over \Lambda^6} {\sigma'\sigma'' \over r},\\
  &&{\cal C}_1 =-6\gamma_0-\frac{4\gamma_1}{\Lambda^3}\left(\pi''+2{\pi' \over r}\right)
  -6 {\gamma_2 \over \Lambda^6}\left({\pi'^2 \over r^2}
    +2{\pi'\pi'' \over r}\right)
  -{24\gamma_3 \over \Lambda^9} {\pi'^2\pi'' \over r^2},\\
  &&{\cal C}_2 =-6\gamma_0 + 4\gamma_1 a
  -\frac{8\gamma_1-12\gamma_2 a}{\Lambda^3} {\pi' \over r}
  - {6\gamma_2- 24\gamma_3 a \over \Lambda^6}{\pi'^2 \over r^2},\\
  &&{\cal C}_3 =-6\gamma_0+4\gamma_1 a
  -\frac{4\gamma_1-6\gamma_2 a}{\Lambda^3}\left(\pi''+{\pi' \over r}\right)
  -{6 \gamma_2 -24\gamma_3 a \over \Lambda^6}{\pi'\pi'' \over r}.
\end{eqnarray}
Here we set $\beta=\xi_2=\xi_4=0$.
One can express the quadratic Lagrangian (\ref{ActionQuadratic})
in a simple form as 
\begin{eqnarray}
  {\cal L}^{(2)}_{\rm DL}&=&
{1 \over 2} (\partial_tQ) J_t (\partial_t Q)
-{1 \over 2} (\partial_rQ) J_r (\partial_r Q)
-{1 \over 2} (\partial_\Omega Q) J_\Omega (\partial_\Omega Q),
\end{eqnarray}
where $Q$ and $J_i$ are the matrices,
\begin{equation}
Q=
\begin{pmatrix}
\phi(t,x) \\
\psi(t,x)
\end{pmatrix}, \quad
J_t=
\begin{pmatrix}
2{\cal A}_1 & {\cal C}_1 \\
{\cal C}_1 & 2{\cal B}_1
\end{pmatrix}, \quad
J_r=
\begin{pmatrix}
2{\cal A}_2 & {\cal C}_2 \\
{\cal C}_2 & 2{\cal B}_2
\end{pmatrix}, \quad
J_\Omega=
\begin{pmatrix}
2{\cal A}_3 & {\cal C}_3 \\
{\cal C}_3 & 2{\cal B}_3
\end{pmatrix}. \quad
\end{equation}
The conditions for avoiding ghost instability are simply given by 
$\det J_t >0$ and $\tr J_t >0$.
These conditions are equivalent to the following explicit conditions,
\begin{eqnarray}
  {\cal A}_1>0, \qquad{\cal B}_1-\frac{{\cal C}_1^2}{4{\cal A}_1 }>0.
\label{conditions:ghostscalar}
\end{eqnarray}
The squared sound speeds in the radial direction $c_r^{2(\pm)}$
are given by the eigenvalues of the matrix, ${\cal M}_r= J_t^{-1} J_r$,
and the squared sound speeds in the angular direction $c_\Omega^{2(\pm)}$
are given by the eigenvalues of the matrix, ${\cal M}_\Omega= J_t^{-1} J_\Omega$.
The condition for avoiding gradient instabilities and superluminal propagation
are given by
\begin{eqnarray}
  0 \leq c_r^{2(\pm)} \leq 1, \qquad   0 \leq c_\Omega^{2(\pm)} \leq 1.
\label{conditions:Cs2scalar}
\end{eqnarray}
In the case of $\lambda={\rm const}$, 
the radial and angular perturbations coincide, $J_r=J_\Omega$. 
Thus we can obtain the dispersion relation 
$\omega=c_{s(\pm)}^{2}k^2$ in Fourier space,
where $c_{s(\pm)}^{2}$  is given by
\begin{eqnarray}
  c_{s(\pm)}^{2}=\frac{-w\pm \sqrt{w^2-4vz}}{2 v},
\end{eqnarray}
with
$v=4{\cal A}_1{\cal B}_1-{\cal C}_1^2$, 
$w=-4({\cal A}_1{\cal B}_2+{\cal A}_2{\cal B}_1)+2{\cal C}_1{\cal C}_2$,
and $z=4{\cal A}_2{\cal B}_2-{\cal C}_2^2$.
Then the conditions for avoiding gradient instability and superluminal propagation
are given by
\begin{eqnarray}
  0 \leq c_{s(\pm)}^{2} \leq 1.
\label{condition2:Cs2}
\end{eqnarray}

\section{Vector Perturbations}\label{sec:vector-perturbations}
In this appendix, we derive the conditions for stability and subluminality for vector perturbations.
Expanding $\pi$ as in (\ref{perturbations}) and integrating out the non-dynamical 
$B$ field, we obtain
\begin{eqnarray}
{\cal L}_{\rm DL}^{(A)}=-C_{tr} F^{tr}F_{tr} -C_{t \theta} F^{t \theta}F_{t \theta}-C_{t \varphi} F^{t \varphi}F_{t \varphi}
-C_{r \theta} F^{r \theta}F_{r \theta}-C_{r \varphi} F^{r \varphi}F_{r \varphi}-C_{ \theta \varphi} F^{\theta \varphi}F_{\theta \varphi},~~~
\label{vectorPL}
\end{eqnarray}
where $F_{\mu\nu}=\partial_\mu A_\nu - \partial_\nu A_\mu$ is the field strength of the vector field and 
\begin{eqnarray}
C_{tr} &=& \frac{1-2\alpha\pi' / \Lambda^3 r}{2(2+a) -2 \pi'' /\Lambda^3},\\
C_{t \theta}&=& C_{t \varphi}=\frac{1-\alpha\pi' / r \Lambda^3-\alpha\pi''/\Lambda^3}{2(2+a)-2\pi'/r \Lambda^3 },\\
C_{r \theta}&=&C_{r \varphi}=\frac{1+a\alpha-\alpha\pi'/r \Lambda^3}{4-2\pi'/ r\Lambda^3-2\pi''/\Lambda^3},\\
C_{\theta \varphi}&=& \frac{1+a\alpha-\alpha\pi''/\Lambda^3}{4(1-\pi'/r\Lambda^3)}.
\end{eqnarray}
One can rewrite this Lagrangian in terms of the electric and magnetic fields,
$E_\mu \equiv  F_{\mu\nu} u ^\nu$ and $B_\mu \equiv \varepsilon_{\mu\nu\alpha\beta}F^{\alpha\beta}u^\nu$,
where $u^\mu$ denotes four velocity; we then have
\begin{eqnarray}
{\cal L}_{\rm DL}^{(A)}=-C_{tr} E^rE_r -C_{t \theta} (E_\theta E^\theta+E_\varphi E^\varphi)
+C_{ \theta \varphi}  B^r B_r+C_{r \theta}(B_\theta B^\theta+B_\varphi B^\varphi)~.
\end{eqnarray}
One can now read off the conditions for the absence of ghosts
\begin{eqnarray}
C_{tr} >0, \qquad C_{r \theta}>0 .
\end{eqnarray}
The sound speeds in various directions of the vector perturbations depend on the polarization modes, 
and are given as follows
\begin{eqnarray}
	c_r^{2(A)}&=&{C_{r\theta} \over C_{t\theta}} \qquad {\rm for~}E_r=B_r=0,\\
	c_{\Omega,1}^{2(A)}&=&{C_{r\theta} \over C_{t r}} \qquad {\rm for~}E_\theta=E_\varphi=B_r=B_\theta=0, \\
	c_{\Omega,2}^{2(A)}&=&{C_{\theta\varphi} \over C_{t\theta}} \qquad {\rm for~} E_r=E_\theta=B_\theta=B_\varphi=0.
\end{eqnarray}
leading to the following conditions
\begin{eqnarray}
0 \leq c_r^{2(A)} \leq 1, \qquad 0 \leq c_{\Omega,1}^{2(A)}\leq 1, \qquad 0 \leq c_{\Omega,2}^{2(A)}\leq 1,
\end{eqnarray}
for avoiding gradient instability and superluminal propagation.

In the $\lambda={\rm const}$ case, 
the vector Lagrangian (\ref{vectorPL}) can be written as 
\begin{eqnarray}
 {\cal L}_{\rm DL}^{(A)}= D_1 F^{0i}F_{0i} + D_2 F^{ij}F_{ij},
\end{eqnarray}
 and 
\begin{eqnarray}
  D_1 = -\frac{1-2\alpha \lambda}{2(2+a -\lambda)}, \qquad
  D_2 = -\frac{1+a\alpha -\alpha\lambda}{8(1 -\lambda)}.
\end{eqnarray}
One can fix the gauge such that $\nabla \cdot {\bf A}=0$,
where ${\bf A}$ is the transverse modes,
and write the vector field as $A_\mu=(0, {\bf A}).$
Then the Lagrangian can be rewrited as 
\begin{eqnarray}
 {\cal L}_{\rm DL}^{(A)}= {\bar D}_1 (\partial_t {\bf A})^2 - {\bar D}_2 (\nabla \times {\bf A})^2,
\end{eqnarray}
where ${\bar D}_1=-D_1$ and ${\bar D}_2=-2D_2$.
Thus the conditions for the absence of ghosts, gradient instability and superluminal 
propagation for the vector modes are given by
\begin{eqnarray}
{\bar D}_1>0, \qquad 0 \leq c_{sA}^{2} \leq 1
\label{conditions:vector}
\end{eqnarray}
$c_{s A}^{2} = {{\bar D}_2 / {\bar D}_1}$.
Note that in the the case, $a= -\lambda$,
the vector Lagrangian is simply given by 
${\cal L}_{\rm DL}^{(A)}=-({\bar D}_1/2) F_{\mu\nu}F^{\mu\nu}$, 
and the sound speed of the vector mode is equal to the speed of light.

\section{Perturbations of the time-dependent solution}\label{sec:perturbation-timedependent}
Here we summarize the kinetic coefficients for perturbations and the sound speeds for all modes on the time-dependent solution given by (\ref{expandedsigma}), (\ref{lambda1}), and (\ref{lambda2}). The coefficients for the solution, corresponding to $\lambda_1=(1+ \sqrt{3})/(2\alpha)$, are given by
\begin{eqnarray}
{\cal A}_1 &=&  3+ c_1\omega^{-1/2} +{\cal O}(\omega^{-1}),\\
{\cal B}_1-{{\cal C}_1^2 \over 4{\cal A}_1}  &=&  {1 \over 2}+ c_2 \omega^{-1}+{\cal  O}(\omega^{-3/2}),\\
c_s^{2(+)}&=&1+c_3\omega^{-1} 	+{\cal  O}(\omega^{-3/2}),\\
c_s^{2(-)}&=&\frac{1}{4}+c_4\omega^{-1/2} +{\cal O}(\omega^{-1}),\\
{\bar D}_1&=& {\bar D}_{1,{\rm decoupled}} + c_5 \omega^{-1/2}+ {\cal O}(\omega^{-1}),\\
c_s^{2(A)}&=& c_{s,{\rm decoupled}}^{2(A)} + c_6 \omega^{-1/2}+ {\cal O}(\omega^{-1}),
\end{eqnarray}
where  ${\bar D}_{1}$ and the sound speed of the vector perturbations in the limit $\omega \to \infty$ are given by 
\begin{eqnarray}
{\bar D}_{1,{\rm decoupled}} &=& \frac{\sqrt{3} \alpha}{2+\sqrt{3}-4\alpha},\\
c_{s,{\rm decoupled}}^{2(A)} &=& \frac{2+\sqrt{3} -4\alpha}{4+4\sqrt{3}-8\alpha}.\\
\label{vectorH}
\end{eqnarray}
Then the conditions for avoiding ghosts, gradient instability, and superluminal propagation of the vector perturbations translate into 
\begin{eqnarray}
0 \leq  \alpha \leq  {1\over 4} (2+\sqrt{3}).
\end{eqnarray}
The next-to-leading order coefficients are given by
\begin{eqnarray}
c_1 &=& \frac{4b \left(\alpha ^3-2 \sqrt{3} \alpha ^2-\sqrt{3} \alpha +\sqrt{3}+2\right) }{\alpha ^2},\\
c_2   &=& -\frac{\left(3 \sqrt{3} \alpha ^3-6\alpha ^2-\left(7+3\sqrt{3}\right) \alpha +3 \sqrt{3}+5\right)^2}{12\alpha ^6},\\
c_3 &=& {2 (2+\sqrt{3}-\sqrt{3} \alpha -2\sqrt{3}  \alpha^2 + \alpha^3 )^2\over 3 \alpha ^6} ,\\
c_4 &=& -\frac{b\left(\alpha ^3-4 \sqrt{3} \alpha ^2+\left(3-2 \sqrt{3}\right) \alpha +2 \sqrt{3}+5\right) }{9\alpha ^2},\\
c_5 &=& -\frac{8b (2 \alpha -1) \left(\sqrt{3}\alpha ^3-6 \alpha ^2-3 \alpha +2\sqrt{3}+3\right) }{9 \left(-4\alpha +\sqrt{3}+2\right)^2 \alpha},\\
c_6 &=&\frac{b(2 \alpha -1) \left(\sqrt{3}\alpha ^3-6 \alpha ^2-3 \alpha +2\sqrt{3}+3\right) }{9 \left(-2\alpha +\sqrt{3}+1\right)^2 \alpha^2}.
\end{eqnarray}

\section{Perturbations of the time-dependent solution in extended theories}\label{sec:perturbation-timedependent-Horn}
In this appendix we focus on one of the time-dependent solutions, corresponding to
$\lambda_{\rm decoupled}=(1 + \sqrt{3})/(2\alpha)$ 
and ${\tilde \lambda}_{\sigma,{\rm decoupled}}= - (1 - \sqrt{3})(2 {\tilde \xi_1})$.
Then, $\lambda_2$ and $\lambda_{\sigma 2}$ can be easily found,
\begin{eqnarray}
  &&\lambda_2 = \frac{2+8\alpha-6\alpha^2-9(2-\sqrt{3})\alpha^3}{9\alpha^3 {\tilde \xi_1}}, \\
  &&\lambda_{\sigma 2} =-\frac{5+3\sqrt{3}-(17+9\sqrt{3})\alpha+(8+6\sqrt{3})\alpha^2+3\alpha^3}{6\alpha^4}.
\end{eqnarray}
and one can evaluate the coefficients of perturbations,
\begin{eqnarray}
  {\cal A}_1 &=& 3 + d_1 \omega^{-1/2}+ {\cal O}(\omega^{-1}),\\
  {\cal B}_1-{{\cal C}_1^2 \over 4{\cal A}_1} &=& 2 + d_2 \omega^{-1/2}+ {\cal O}(\omega^{-1}),\\
 c_s^{2(+)} &=& {1 \over 4} + d_3 \omega^{-1/2}+ {\cal O}(\omega^{-1}),\\
   c_s^{2(-)} &=& {1 \over 4} + d_4 \omega^{-1/2}+ {\cal O}(\omega^{-1}),\\
 {\bar D}_1&=& {\bar D}_{1,{\rm decoupled}} + d_5 \omega^{-1/2}+ {\cal O}(\omega^{-1}),\\
 c_s^{2(A)}&=& c_{s,{\rm decoupled}}^{2(A)} + d_6 \omega^{-1/2}+ {\cal O}(\omega^{-1}),
\end{eqnarray}
where ${\bar D}_1$ and $ c_s^{2(A)}$ are the same as in (\ref{vectorH}), and 
\begin{eqnarray}
  d_1 &=& \frac{3+5\sqrt{3}-(9-11\sqrt{3})\alpha -18\alpha^2+3(8-5\sqrt{3})\alpha^3}{2 \alpha^2 {\tilde \xi}_1},\\
  d_2   &=& -\frac{9+5\sqrt{3}-(27+17\sqrt{3})\alpha +2(9+4\sqrt{3})\alpha^2
+3\sqrt{3}\alpha^3}{2 \alpha^4 } {\tilde \xi}_1,\\
 d_3 &=& {d_1 \over 24},\\
   d_4 &=& \frac{-9-19\sqrt{3}+(27-13\sqrt{3})\alpha 
-6(3-4\sqrt{3})\alpha^2-3\sqrt{3}\alpha^3}{72 \alpha^2 {\tilde \xi}_1},\\
 d_5 &=& \frac{4(1-2\alpha)\left(2+8\alpha-6\alpha^2-9(2-\sqrt{3})\alpha^3\right)}
{9(2+\sqrt{3}-4\alpha)^2 \alpha {\tilde \xi}_1},\\
 d_6 &=& -\frac{(1-2\alpha)\left(2+8\alpha-6\alpha^2-9(2-\sqrt{3})\alpha^3\right)}
{18(1+\sqrt{3}-2\alpha)^2 \alpha^2 {\tilde \xi}_1}.
\end{eqnarray}

\bibliographystyle{JHEPmodplain}
\bibliography{references}

\end{document}